\documentclass[onecolumn,superscriptaddress,amsmath,amssymb,floatfix,aps,notitlepage,prx,longbibliography]{revtex4-1}
\usepackage{mathrsfs, hyperref}
\usepackage{amssymb, amsbsy, amsmath, latexsym, dsfont, array, layout, graphics,mathrsfs,braket,amsfonts,amsthm,
  amssymb,graphicx,subfigure, youngtab,color,bm,mathtools,braket,verbatim,url,cleveref,natbib,hypernat,extarrows,inputenc,multirow}
\usepackage[usernames,dvipsnames]{xcolor}

\newcommand{\bs}[1]{\boldsymbol{#1}}
\newcommand{\mbf}[1]{\mathbf{#1}}

\definecolor{AtomicTangerine}{rgb}{1.0, 0.6, 0.4}

\newcommand{\externalStress}[1]{{\color{red}#1}}
\newcommand{\inclusionI}[1]{{\color{blue}#1}}
\newcommand{\inclusionJ}[1]{{\color{OliveGreen}#1}}

\makeatletter
\newcommand*{\inlineequation}[2][]{%
  \begingroup
    \refstepcounter{equation}%
    \ifx\\#1\\%
    \else
      \label{#1}%
    \fi
    \relpenalty=10000 %
    \binoppenalty=10000 %
    \ensuremath{%
      #2%
    }%
    ~\@eqnnum
  \endgroup
}
\makeatother

\newcolumntype{M}[1]{>{\centering\arraybackslash}m{#1}}
\newcolumntype{N}{@{}m{0pt}@{}}

\allowdisplaybreaks

\newcommand{\splitatcommas}[1]{%
  \begingroup
  \begingroup\lccode`~=`, \lowercase{\endgroup
    \edef~{\mathchar\the\mathcode`, \penalty0 \noexpand\hspace{0pt plus 1em}}%
  }\mathcode`,="8000 #1%
  \endgroup
}

\begin{document}
\title{Elastic multipole method for describing linear deformation of\\infinite 2D solid structures with circular holes and inclusions}
\author{Siddhartha Sarkar}
\affiliation{Department of Electrical Engineering, Princeton University, Princeton, NJ 08544, USA}
\author{Matja\v{z} \v{C}ebron}
\affiliation{Faculty of Mechanical Engineering, University of Ljubljana, SI-1000 Ljubljana, Slovenia}
\author{Miha Brojan}
\email{miha.brojan@fs.uni-lj.si}
\affiliation{Faculty of Mechanical Engineering, University of Ljubljana, SI-1000 Ljubljana, Slovenia}
\author{Andrej Ko\v{s}mrlj}
\email{andrej@princeton.edu}
\affiliation{Department of Mechanical and Aerospace Engineering, Princeton University, Princeton, NJ 08544, USA}
\affiliation{Princeton Institute for the Science and Technology of Materials, Princeton University, Princeton, NJ 08544, USA}

\begin{abstract}
Elastic materials with holes and inclusions are important in a large variety of contexts ranging from construction material to biological membranes. More recently, they have also been exploited in mechanical metamaterials, where the geometry of highly deformable structures is responsible for their unusual properties, such as negative Poisson's ratio, mechanical cloaking, and tunable phononic band gaps. Understanding how such structures deform in response to applied external loads is thus crucial for designing novel mechanical metamaterials. Here we present a method for predicting the linear response of infinite 2D solid structures with circular holes and inclusions by employing analogies with electrostatics. Just like an external electric field induces polarization (dipoles, quadrupoles and other multipoles) of conductive and dielectric objects,  external stress induces elastic multipoles inside holes and inclusions. Stresses generated by these induced elastic multipoles then lead to interactions between holes and inclusions, which induce additional polarization and thus additional deformation of holes and inclusions. We present a method that expands the induced polarization in a series of elastic multipoles, which systematically takes into account the interactions of inclusions and holes with the external stress field and also between them. The results of our method show good agreement with both linear finite element simulations and experiments.
\end{abstract}

\maketitle

\section{Introduction}
Elastic materials with holes and inclusions have been studied extensively in materials science. Typically, the goal is to homogenize the microscale distribution of holes and inclusions to obtain effective material properties on the macroscale~\cite{Eshelby,Hashin,Castaneda,Torquato}, where the detailed micropattern of deformations and stresses is ignored. On the other hand, it has recently been recognized that the microscale interactions between proteins embedded in biological membranes can promote the assembly of ordered protein structures~\cite{ProteinAssembly2,ProteinAssembly,Haselwandter} and can also facilitate the entry of virus particles into cells~\cite{Protein}.
Furthermore, in mechanical metamaterials~\cite{bertoldi2017flexible}, the geometry, topology and contrasting elastic properties of different materials are exploited to achieve extraordinary functionalities, such as shape morphing~\cite{ShapeMorph,ShapeMorph2}, mechanical cloaking~\cite{Cloaking,Cloaking2,Cloaking3}, negative Poisson's ratio~\cite{almgren1985isotropic,lakes1987foam,Auxetic2,Auxetic,babaee20133d}, negative thermal expansion~\cite{NegativeThermalExpansion,NegativeThermalExpansion2}, effective negative swelling~\cite{NegativeSwelling,NegativeSwelling2,NegativeSwelling3}, and tunable phononic band gaps~\cite{BandGap3,BandGap,BandGap2}. At the heart of these functionalities are deformation patterns of such materials with holes and inclusions. Therefore, understanding how these structures deform under applied external load is crucial for the design of novel metamaterials.

\begin{figure}[t!]
\centering
\includegraphics[scale=1]{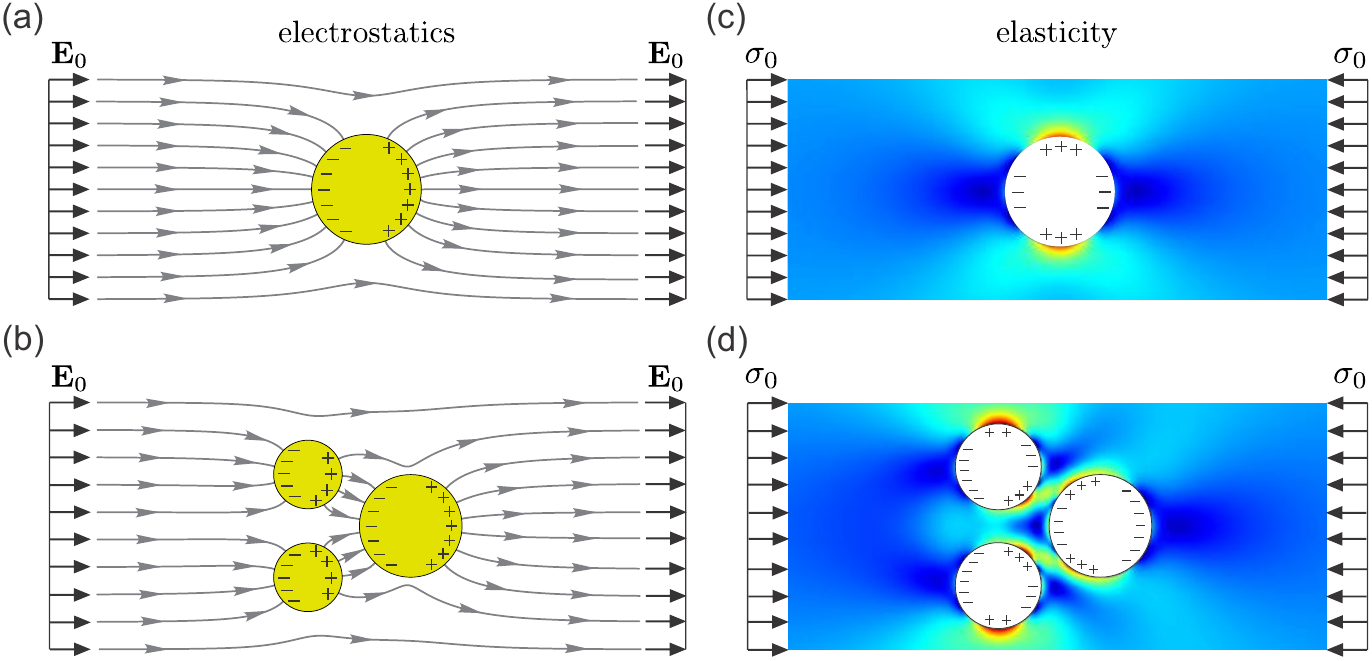}
\caption{
Induction in electrostatics and elasticity. (a,b) 
Induced polarization due to the external electric field $\mathbf{E}_0$ of (a)~a single and (b)~multiple conducting spheres (yellow). The resultant electric field lines are shown in grey color.
(c,d)~Induced quadrupoles due to external uniaxial compressive stress $\sigma_0$ in (c)~a single and (d)~multiple circular holes (white disks) embedded in an elastic matrix. Heat maps show the von Mises stress field, where red and blue colors indicate regions of high and low von Mises stress, respectively.}
     \label{fig:induction}
\end{figure}

Linear deformations of infinite thin plates with circular holes under external load have been studied extensively over the years~\cite{Green,Haddon,Ukadgaonker,Ting,Hoang08}. The solution for one hole can be easily obtained using standard techniques~\cite{Barber} and the solution for two holes can be constructed with conformal maps and complex analysis~\cite{Haddon}. Green demonstrated how to construct a solution for infinite thin plates with any number of holes~\cite{Green} by expanding the Airy stress function around each hole in terms of the Michell solution for biharmonic functions~\cite{Michell}. However, it remained unclear how this procedure could be generalized to finite structures with boundaries.

Deformations of thin membranes with infinitely rigid inclusions have also received a lot of attention, especially in the context of rigid proteins embedded in biological membranes~\cite{goulian1993long,park1996interactions,golestanian1996fluctuationPRE,golestanian1996fluctuationEPL,weikl1998interaction,ProteinAssembly,yolcu2012membrane,Deserno,Haselwandter,Protein,Purohit}. Several different approaches were developed to study the elastic and entropic interactions between inclusions, such as multipole expansion~\cite{goulian1993long}, the effective field theory approach~\cite{yolcu2012membrane,Deserno}, and homogenization~\cite{Protein}. Even though these articles considered membrane bending, the governing equation for the out-of-plane displacement is also biharmonic to the lowest order. Hence these methods could be adapted to investigate the in-plane deformations of plates with rigid inclusions.

In two companion papers, we have generalized Green's method~\cite{Green} by employing analogies with electrostatics to describe the linear response of a thin elastic plate (plane stress) or an infinitely thick elastic matrix (plane strain) with embedded cylindrical holes and inclusions, which can be treated as a 2D problem with circular holes and inclusions.
 Just like a polarized conductive object in an external electric field can be described by induced dipole (see Fig.~\ref{fig:induction}a), a hole deformed by the external load can be described by induced elastic quadrupoles (see Fig.~\ref{fig:induction}c). Circular inclusions in the elastic matrix under external load are analogous to dielectric objects in an external electric field. When multiple conductive objects are placed in an external electric field, the induced polarizations generate additional electric fields, which lead to further charge redistribution on the surface of conductive objects (see Fig.~\ref{fig:induction}b). Similarly, induced quadrupoles in deformed holes generate additional stresses in the elastic matrix, which lead to further deformations of holes (see Fig.~\ref{fig:induction}d).

In this paper, we present a method to describe linear deformations of circular holes and inclusions embedded in an \textit{infinite} 2D elastic matrix under small external loads by systematically expanding induced polarization of each hole/inclusion in terms of elastic multipoles that are related to terms in the Michell solution for biharmonic functions~\cite{Michell}. The results of this method are compared with linear finite element simulations and experiments. We show that the error decreases exponentially as the maximum degree of elastic multipoles is increased. In the companion paper~\cite{sarkar2020image}, we describe how this method can be generalized to \textit{finite size} structures by employing ideas of image charges, which become important for holes and inclusions near boundaries.

The remaining part of the paper is organized as follows. In Section~\ref{sec:analogy}, we review the analogy between electrostatics and 2D linear elasticity and introduce important concepts borrowed from electrostatics. In Section~\ref{sec:ElasticMultipoleMethod}, we describe the method for evaluating linear deformation of structures with holes and inclusions under external load, which is compared with linear finite element simulations and experiments.
In Section~\ref{sec:Conclusion}, we give concluding remarks and comment on the extensions of this method to the nonlinear deformation regime, which is also important for the analysis of mechanical metamaterials.

\section{Analogy between electrostatics and 2D linear elasticity}\label{sec:analogy}

The analogy between electrostatics and 2D linear elasticity can be recognized, when the governing equations are formulated in terms of the electric potential $U$~\cite{Jackson} and the Airy stress function $\chi$~\cite{Barber}, respectively, which are summarized in Table~\ref{tab:ElecVsElas}.  The measurable fields, namely the electric field $\mathbf{E}$ and the stress tensor field $\sigma_{ij}$, are obtained by taking spatial derivatives of these scalar functions, as shown in Eqs.~(\ref{eq:ElectricPotential}) and~(\ref{eq:AiryStress}), where $\epsilon_{ij}$ is the permutation symbol ($\epsilon_{12}=-\epsilon_{21}=1, \epsilon_{11}=\epsilon_{22}=0$) and summation over repeated indices is implied. The most compelling aspect of the formulations in terms of scalar functions $U$ and $\chi$ is that  Faraday's law in electrostatics in Eq.~(\ref{eq:Conservative}) and the force balance in elasticity in Eq.~(\ref{eq:Equilibrium}) are automatically satisfied. Moreover, the governing equations for these scalar functions take simple forms as shown in Eqs.~(\ref{eq:GovElectrostatics}) and~(\ref{eq:GovElasticity}).
 Equation~(\ref{eq:GovElectrostatics}) describes the well known Gauss's law, where $\rho_e$ is the electric charge density and $\epsilon_e$ is the permittivity of the material. The analogous Eq.~(\ref{eq:GovElasticity}) in elasticity describes the (in)compatibility conditions~\cite{Chaikin,moshe2014plane}, where $E$ is the 2D Young's modulus and $\rho$ is the elastic charge density associated with defects, which are sources of incompatibility. In the absence of electric charges ($\rho_e=0$) the electric potential $U$ is a harmonic function (see Eq.~(\ref{eq:GovElectrostatics})), while in the absence of defects ($\rho=0$) the Airy stress function $\chi$ is a biharmonic function (see Eq.~(\ref{eq:GovElasticity})). 
\begin{table}[t!]
    \caption{Comparison between equations in electrostatics and 2D linear elasticity}
    \label{tab:ElecVsElas}
    \centering
    \begin{tabular}{|M{5cm}|M{5cm}|M{5cm}|N}
        \hline
                           & {\bf Electrostatics} & {\bf Elasticity} & \\[10pt]
        \hline
         Scalar potentials & $U$ &  $\chi$ & \\[10pt]
        \hline
        Fields & \inlineequation[eq:ElectricPotential]{\mathbf{E} = -\boldsymbol{\nabla}U} & \inlineequation[eq:AiryStress]{\sigma_{ij}=\epsilon_{ik}\epsilon_{jl}\frac{\partial^2 \chi}{\partial x_k \partial x_l}} & \\[10pt]
        \hline
         Properties of scalar functions & \inlineequation[eq:Conservative]{\boldsymbol{\nabla}\times \mathbf{E}=-\boldsymbol{\nabla}\times \boldsymbol{\nabla}U=\mathbf{0}} & \inlineequation[eq:Equilibrium]{\frac{\partial \sigma_{ij}}{\partial x_j} = \epsilon_{ik}\epsilon_{jl}\frac{\partial}{\partial x_j}\frac{\partial^2 \chi}{\partial x_k \partial x_l} = 0} & \\[10pt]
        \hline
         Governing equations & \inlineequation[eq:GovElectrostatics]{\Delta U = -\rho_e/\epsilon_e} & \inlineequation[eq:GovElasticity]{\Delta \Delta \chi = E\rho} & \\[10pt]
        \hline
    \end{tabular}
\end{table}

When a conductive object is placed in an external electric field, it gets polarized due to the redistribution of charges (see Fig.~\ref{fig:induction}a). This induced polarization generates an additional electric field outside the conductive object, which can be expanded in terms of fictitious multipoles (dipole, quadrupole, and other multipoles) located at the center of the conductive object~\cite{Jackson}. Note that the induced polarization does not include a monopole charge, because the total topological charge is  conserved~\cite{Jackson}. Similarly, a hole or inclusion embedded in an elastic matrix gets polarized when the external load is applied (see Fig.~\ref{fig:induction}c). The additional stresses in the elastic matrix due to this induced polarization can again be expanded in terms of fictitious \textit{elastic multipoles} (quadrupoles and other multipoles) located at the center of hole/inclusion. In elasticity, the induced polarization does not include disclinations (topological monopole) and dislocations (topological dipole), which are topological defects~\cite{Chaikin}. In order to demonstrate this, we first briefly present the multipoles and induction in electrostatics and describe the meaning of their counterparts in elasticity.

\begin{figure}[t!]  \includegraphics[scale=1]{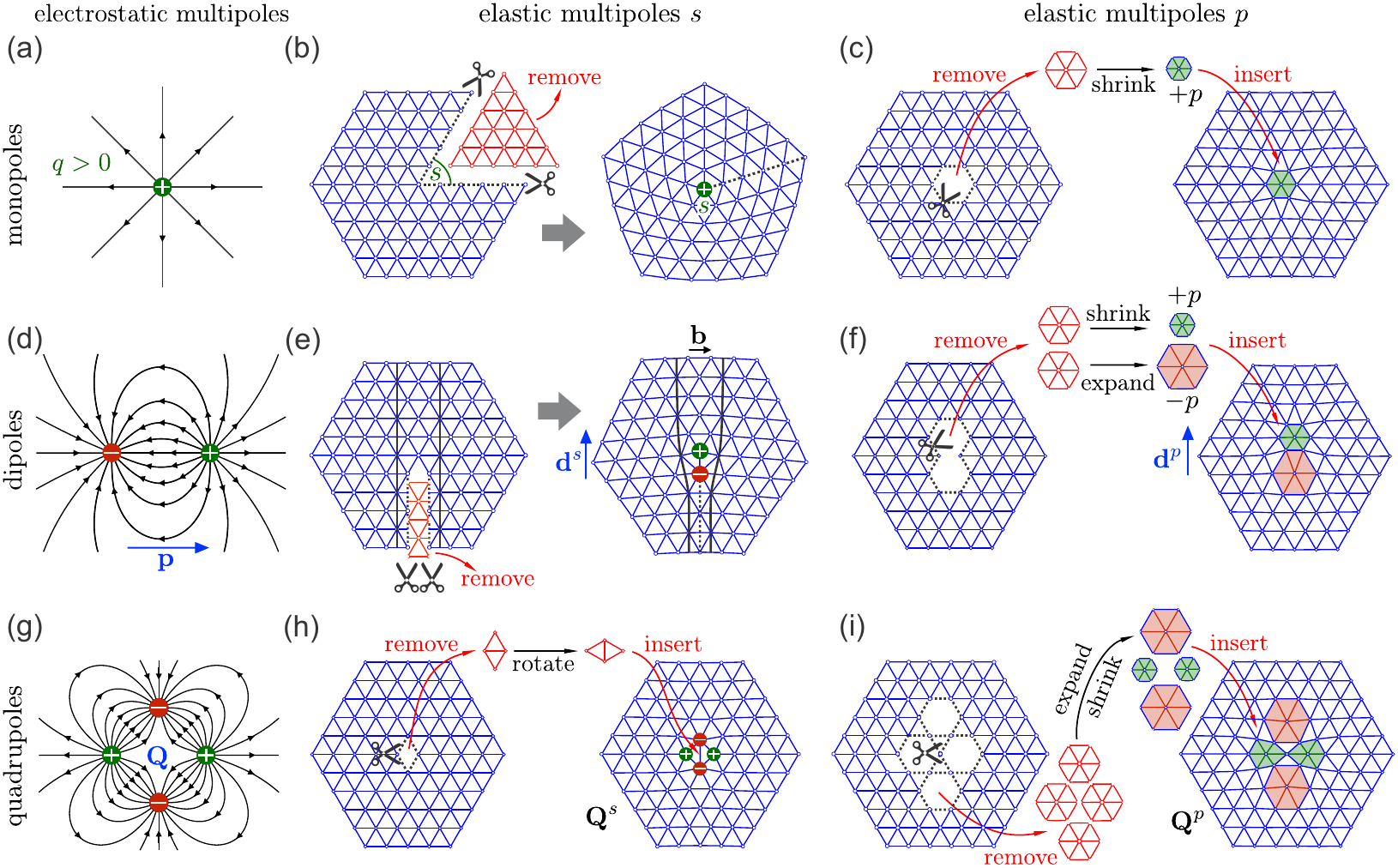}
  \caption{Multipoles in electrostatics and 2D elasticity. (a)~An electrostatic monopole with positive charge $q$ (green) generates an outward radial electric field (black lines). For a monopole with negative charge, the direction of  the electric field is reversed. (b)~In 2D elasticity a disclination defect (topological monopole) with charge $s$ forms upon removal ($s>0$) or insertion ($s<0$) of a wedge of material (red), where $|s|$ is the wedge angle.
  (c)~In 2D elasticity, a non-topological monopole $p>0$ ($p<0$) corresponds to a local isotropic contraction (expansion) of the material. (d)~An electrostatic dipole $\mathbf{p}$ is formed when a positive (green) and a negative (red) charge of equal magnitude are brought close together. The resulting electric field lines are shown with black lines. (e)~In 2D elasticity, a dislocation (topological dipole) forms upon removal or insertion of a semi-infinite strip of material of width $|\mathbf{b}|$ and is represented by the Burgers vector $\mathbf{b}$. In a triangular lattice, the dislocation corresponds to two adjacent disclinations of opposite charges. The two black lines indicate the positions of points before and after the removal of a semi-infinite strip (red) from crystal. A dipole moment $\mathbf{d}^s$ can be defined in the direction from negative to positive disclination and its magnitude is equal to the distance between two disclinations times the magnitude of charge of each disclination. (f)~In 2D elasticity, a non-topological dipole $\mathbf{d}^p$ is formed when a positive (green) and a negative (red) non-topological charge of equal magnitude are brought close together. (g)~An electrostatic quadrupole $\mbf{Q}$ consisting of four charges at the vertices of a square with opposite charges at the adjacent vertices. The resulting electric field lines are shown with black lines. (h)~In 2D elasticity, a quadrupole $\mbf{Q}^s$ is represented by four disclinations at the vertices of a square with opposite charges at the adjacent vertices. Due to the quadrupole $\mbf{Q}^s$,  material locally expands in the direction of  positive disclinations and locally contracts in the direction of negative disclinations, while the total area remains locally unchanged. (i)~In 2D elasticity, a quadrupole $\mbf{Q}^p$ is represented by four non-topological monopoles at the vertices of a square with opposite charges at the adjacent vertices. }
  \label{Fig:Multipoles}
\end{figure}

\subsection{Monopoles}
In electrostatics, a \textit{topological monopole} is defined as the electric charge density distribution
proportional to the Dirac delta function, i.e. $\rho_e = q \delta(\mbf{x}-\mbf{x}_0)$, where $q$ is the charge and $\mbf{x}_0$ denotes its position. The electric potential $U_m(\mbf{x}-\mbf{x}_0|q)$ in 2D is then obtained by solving the governing equation as~\cite{Jackson}
\begin{equation}
  \label{eq:ElecMonopole}
  \Delta U_m =
  -\frac{q}{\epsilon_e}\delta(\mbf{x}-\mbf{x}_0)\quad\Longrightarrow\quad
  U_m(\mbf{x}-\mbf{x}_0|q) = -\frac{q}{2\pi \epsilon_e}\ln  |\mbf{x}-\mbf{x}_0|.
\end{equation}
For the positive monopole charge, the electric field $\mbf{E}_m=-\bs{\nabla} U_m$ is pointing radially outward (see Fig.~\ref{Fig:Multipoles}a). Note that the total charge is topologically conserved~\cite{Jackson}.

Similarly, we can define a \textit{topological monopole} in 2D elasticity as the charge density proportional to the Dirac delta function, i.e. $\rho = s \delta(\boldsymbol{x}-\boldsymbol{x}_0)$, where $s$ is the charge and $\mbf{x}_0$ denotes its position. Topological monopoles are called disclinations and their Airy stress function $\chi_m^s(\mbf{x}-\mbf{x}_0|s)$ can be obtained by solving the governing equation as~\cite{Chaikin,Moshe2}
\begin{equation}
  \label{eq:ElasMonopole}
  \Delta \Delta \chi_m^s = E s \delta (\mbf{x}-\mbf{x}_0) \quad \Longrightarrow \quad \chi_m^s(\mbf{x}-\mbf{x}_0|s)  = \frac{E s}{8\pi}|\mbf{x}-\mbf{x}_0|^2 \big(\ln|\mbf{x}-\mbf{x}_0| -1/2\big).
\end{equation}
The physical interpretation of topological monopoles in 2D elasticity comes from condensed matter theory. When a wedge with angle $s$ is cut out from a 2D elastic material and the newly created boundaries of the remaining material are glued together, a positive disclination defect of charge $s$ is formed (see Fig.~\ref{Fig:Multipoles}b). The negative disclination with charge $s<0$ corresponds to the insertion of a wedge with angle $|s|$.
The stresses generated by these operations are described with the Airy stress function in Eq.~(\ref{eq:ElasMonopole}).~\cite{Chaikin,Moshe2}

Unlike in electrostatics, we can also define a \textit{non-topological monopole} in 2D elasticity as the charge density proportional to $\rho = p \Delta_0 \delta(\boldsymbol{x}-\boldsymbol{x}_0)$, where $p$ is the charge, $\mbf{x}_0$ denotes its position, and $\Delta_0$ corresponds to the Laplace operator with respect to $\mbf{x}_0$. The corresponding Airy stress function $\chi_m^p(\mbf{x}-\mbf{x}_0|p)$ can be obtained by solving the governing equation as~\cite{Moshe2}
\begin{equation}
  \label{eq:ElasMonopoleP}
  \Delta \Delta \chi_m^p = E p \Delta_0\delta (\mbf{x}-\mbf{x}_0) \quad \Longrightarrow \quad \chi_m^p(\mbf{x}-\mbf{x}_0|p)  = \frac{E p}{2 \pi} (\ln(|\mbf{x}-\mbf{x}_0|)+1/2).
\end{equation}
Note that the Airy stress functions for the non-topological monopole $p$ and for the topological monopole $s$ are related via $\chi_m^p(\mbf{x}-\mbf{x}_0|p) = \Delta_0 \chi_m^s(\mbf{x}-\mbf{x}_0|p)$. The constant term in Eq.~(\ref{eq:ElasMonopoleP}) does not generate any stresses and can thus be omitted. 
A positive (negative) non-topological monopole with charge $p>0$ ($p<0$) is related to a local isotropic contraction (expansion) of the material (see Fig~\ref{Fig:Multipoles}c).~\cite{Moshe1,Moshe2}

\subsection{Dipoles}
An electrostatic dipole is formed at $\mbf{x}_0$ when two opposite charges $\pm q$ are located at $\mbf{x}_\pm=\mbf{x}_0 \pm \mbf{a}/2$ (see Fig.~\ref{Fig:Multipoles}d). The electric potential for a dipole in 2D is thus
\begin{equation}
  \label{eq:ElecDipole}
  U_d(\mbf{x}-\mbf{x}_0|\mbf{p}) = U_m(\mbf{x}-\mbf{x}_+|q)+ U_m(\mbf{x}-\mbf{x}_-|-q) \ \xlongrightarrow{{|\mbf{a}|\to 0}}\ 
  \frac{\mathbf{p}\cdot (\mbf{x}-\mbf{x_0})}{2\pi \epsilon_e |\mbf{x}-\mbf{x}_0|^2},
\end{equation}
where we introduced the dipole moment $\mbf{p}=q \mbf{a}$.~\cite{Jackson} Note that in electrostatics dipoles and all higher-order multipoles are \textit{non-topological}~\cite{Jackson}.

Similarly, a dipole $\mbf{d}^s=s \mbf{a}$ in 2D elasticity  is formed when two disclination defects of opposite charges $\pm s$ are located at $\mbf{x}_\pm=\mbf{x}_0 \pm \mbf{a}/2$ (see Fig.~\ref{Fig:Multipoles}e). Dipoles are called dislocations and their Airy stress function is~\cite{Chaikin}
\begin{equation}
  \label{eq:ElasDipole}
  \chi^s_d(\mbf{x}-\mbf{x}_0|\mbf{d}^s) = \chi_m^s(\mbf{x}-\mbf{x}_+|s)+ \chi_m^s(\mbf{x}-\mbf{x}_-|-s)\ \xlongrightarrow{{|\mbf{a}|\to 0}}\ 
  -\frac{E}{4\pi} \mbf{d^s}\cdot
  (\mbf{x}-\mbf{x}_0) \ln |\mbf{x}-\mbf{x}_0|.
\end{equation}
Dislocation is a \textit{topological} defect, which forms upon removal or insertion of a semi-infinite strip of material of width $|\mathbf{b}|$ (see Fig.~\ref{Fig:Multipoles}e). Note that dislocations are conventionally represented by the Burgers vector $\mathbf{b}$, which is equal to the dipole moment $\mbf{d}^s$ rotated by 90$^\circ$, i.e. $b_i = \epsilon_{ij} d_j^s$.~\cite{Chaikin,Moshe2}

In 2D elasticity, we can define another \textit{non-topological} dipole $\mbf{d}^p=p \mbf{a}$, which is formed when two non-topological monopoles of opposite charges $\pm p$ are located at $\mbf{x}_\pm=\mbf{x}_0 \pm \mbf{a}/2$ (see Fig.~\ref{Fig:Multipoles}f). Their Airy stress function is
\begin{equation}
  \label{eq:ElasDipole}
  \chi^p_d(\mbf{x}-\mbf{x}_0|\mbf{d}^p) = \chi_m^p(\mbf{x}-\mbf{x}_+|p)+ \chi_m^p(\mbf{x}-\mbf{x}_-|-p)\ \xlongrightarrow{{|\mbf{a}|\to 0}}\ 
   -\frac{E}{2 \pi}\frac{\mbf{d}^p \cdot (\mbf{x}-\mbf{x}_0)}{|\mbf{x}-\mbf{x}_0|^2}.
\end{equation}

\subsection{Quadrupoles}
An electrostatic quadrupole $\mbf{Q}$ in 2D is formed when two positive and negative charges are placed symmetrically around  $\mbf{x}_0$, such that charges $q_i=q(-1)^i$ are placed at positions $\mbf{x}_i=\mbf{x}_0+a \big(\cos(\theta+i\pi/2),\sin(\theta+i\pi/2)\big)$, where $i\in \{0,1,2,3\}$ and angle $\theta$ describes the orientation of quadrupole (see Fig.~\ref{Fig:Multipoles}g). The electric potential of the quadrupole is thus
\begin{equation}
  \label{eq:ElecQuadrupole}
  U_Q(r,\varphi|Q,\theta) =  \sum_{i=0}^{3}  U_m(\mbf{x}-\mbf{x}_i| q_i) \ \xlongrightarrow{{a\to 0}}\  \frac{Q \cos\big(2 (\varphi-\theta)\big)}{\pi \epsilon_e  r^2},
\end{equation}
where we introduced the quadrupole moment $Q=qa^2$ and polar coordinates ($r=\sqrt{(x-x_0)^2+(y-y_0)^2}$, $\varphi=\arctan[(y-y_0)/(x-x_0)]$) centered at $\mbf{x}_0$.

Similarly, an elastic quadrupole $\mbf{Q}^s$ is formed when two positive and negative disclinations are placed symmetrically around $\mbf{x}_0$, such that disclinations with charges $s_i=s(-1)^i$ are placed at positions $\mbf{x}_i=\mbf{x}_0+a \big(\cos(\theta+i\pi/2),\sin(\theta+i\pi/2)\big)$, where $i\in \{0,1,2,3\}$ and angle $\theta$ describes the orientation of quadrupole (see Fig.~\ref{Fig:Multipoles}h). The Airy stress function for  quadrupole $\mbf{Q}^s$ in polar coordinates is thus 
\begin{equation}
  \label{eq:ElasQuadrupoleS}
  \chi_Q^s(r,\varphi|Q^s,\theta) =  \sum_{i=0}^{3}  \chi_m^s(\mbf{x}-\mbf{x}_i|s_i)  \ \xlongrightarrow{{a\to 0}}\  \frac{E Q^s \cos\big(2 (\varphi-\theta)\big)}{4 \pi},
\end{equation}
where we introduced the quadrupole moment $Q^s=s a^2$. The elastic quadrupole $\mbf{Q}^s$ causes the material to locally expand in the $\theta$ direction and locally contract in the orthogonal direction (see Fig.~\ref{Fig:Multipoles}h). Note that the quadrupole $\mbf{Q}^s$ is \textit{non-topological}~\cite{Moshe1,Moshe2}. 

In elasticity, we can define another quadrupole $\mbf{Q}^p$, which is formed when two positive and negative non-topological monopoles are placed symmetrically around $\mbf{x}_0$, such that non-topological charges $p_i=p(-1)^i$ are placed at positions $\mbf{x}_i=\mbf{x}_0+a \big(\cos(\theta+i\pi/2),\sin(\theta+i\pi/2)\big)$, where $i\in \{0,1,2,3\}$ and angle $\theta$ describes the orientation of quadrupole (see Fig.~\ref{Fig:Multipoles}i). The Airy stress function for  quadrupole $\mbf{Q}^p$ in polar coordinates is thus 
\begin{equation}
  \label{eq:ElasQuadrupoleP}
  \chi_Q^p(r,\varphi|Q^p,\theta) =  \sum_{i=0}^{3}  \chi_m^p(\mbf{x}-\mbf{x}_i|p_i)  \ \xlongrightarrow{{a\to 0}}\  \frac{E Q^p \cos\big(2 (\varphi-\theta)\big)}{\pi r^2},
\end{equation}
where we introduced the quadrupole moment $Q^p=p a^2$.

\subsection{Higher-order multipoles} 
The procedure described in the previous sections can be generalized to define higher-order multipoles $\mbf{Q}^s_n$ and $\mbf{Q}^p_n$. In 2D the quadrupole $\mbf{Q}^s$ is generalized by placing $n$ positive and $n$ negative disclinations symmetrically around $\mbf{x}_0$, such that disclinations of charges $s_i=s(-1)^i$ are placed at positions $\mbf{x}_i=\mbf{x}_0+a \big(\cos(\theta+i\pi/n),\sin(\theta+i\pi /n)\big)$, where $i~\in~\{0,1,\ldots,2n-1\}$ and angle $\theta$ describes the orientation of multipole. The Airy stress functions for such multipoles $\mbf{Q}^s_n$ in polar coordinates are
\begin{equation}
  \label{eq:ElasMultipoleQ}
 \chi_n^s(r,\varphi|Q_n^s,\theta) =\sum_{i=0}^{2n-1}  \chi_m^s(\mbf{x}-\mbf{x}_i|s_i) \ \xlongrightarrow{{a\to 0}}\  \frac{E Q_n^s \cos\big(n (\varphi-\theta)\big)}{4(n-1) \pi r^{n-2}},
\end{equation}
where we introduced the multipole moment $Q_n^s=sa^n$.

The quadrupole $\mbf{Q}^p$ is generalized to higher-order multipoles by placing $n$ positive and $n$ negative non-topological monopoles symmetrically around $\mbf{x}_0$, such that charges of strength $p_i=p(-1)^i$ are placed at positions $\mbf{x}_i=\mbf{x}_0+a \big(\cos(\theta+i\pi/n),\sin(\theta+i\pi /n)\big)$, where $i~\in~\{0,1,\ldots,2n-1\}$ and angle $\theta$ describes the orientation of multipole. The Airy stress functions for such multipoles $\mbf{Q}^p_n$ in polar coordinates are
\begin{equation}
  \label{eq:ElasMultipoleP}
\chi_n^p(r,\varphi|Q_n^p,\theta) =\sum_{i=0}^{2n-1}  \chi_m^p(\mbf{x}-\mbf{x}_i|p_i) \ \xlongrightarrow{{a\to 0}}\ -
\frac{E Q_n^p \cos\big(n (\varphi-\theta)\big)}{\pi r^{n}},
\end{equation}
where we introduced the multipole moment $Q_n^p=pa^n$.

\subsection{Multipoles vs. the Michell solution for biharmonic functions} \label{sec:Michell}
The elastic multipoles of types $s$ and $p$ introduced in the previous sections are closely related to the general solution of the biharmonic equation $\Delta \Delta \chi = 0$, due to Michell~\cite{Michell}, which is given in polar coordinates $(r,\varphi)$ as
\begin{equation}
  \begin{split}
  \chi(r,\varphi) &=  A_0 r^2 + B_0 r^2 \ln r + C_0 \ln r + I \varphi \\
   &\phantom{=} +(A_1 r+ B_1 r^{-1}+B_1' r \varphi + C_1 r^3 +  D_1 r \ln r) \cos\varphi\\
   &\phantom{=} +(E_1 r+ F_1 r^{-1}+F_1' r \varphi + G_1 r^3
     +  H_1 r \ln
     r)\sin\varphi\\
   &\phantom{=} +\sum_{n=2}^\infty (A_n r^n+ B_n r^{-n}+ C_nr^{n+2}
   + D_{n}
   r^{-n+2})\cos (n\varphi)\\
   &\phantom{=}+\sum_{n=2}^\infty (E_n r^n+ F_n r^{-n}+ G_n r^{n+2} + H_{n} r^{-n+2})\sin (n\varphi).
  \end{split}
  \label{eq:Michell}
\end{equation}
The Michell solution above contains the Airy stress functions corresponding
to  multipoles located at the origin: disclination ({$r^2\ln r $}), dislocation ({$r\ln r \cos\varphi$}, {$r\ln r \sin\varphi$}), non-topological monopole ({$\ln r $}), non-topological dipole ($\cos \varphi/r$, $\sin \varphi/r$), quadrupole $\mbf{Q}^s$ ({$\cos (2\varphi)$}, {$\sin (2\varphi)$}), quadrupole $\mbf{Q}^p$ ({$\cos (2\varphi)/r^2$}, {$\sin (2\varphi)/r^2$}), as well as all higher-order multipoles $\mbf{Q}^s_n$ and $\mbf{Q}^p_n$ (see Eqs.~(\ref{eq:ElasMultipoleQ}, \ref{eq:ElasMultipoleP})). Note that the Michell solution also contains terms that increase faster than $r^2$ far away from the origin. These terms are associated with stresses that increase away from the origin and can be interpreted as multipoles located at infinity~\cite{Moshe6}. Due to the connection with elastic multipoles we refer to coefficients $A_i, B_i, \dots, H_i$ in the Michell solution as the amplitudes of multipoles.

\subsection{Induction}\label{sec:induction}
As mentioned previously, the external electric field induces polarization in conducting and dielectric objects. Similarly, external stress induces elastic quadrupoles inside holes and inclusions. To make this analogy concrete, we first demonstrate how external electric field in 2D polarizes a single conductive or dielectric disk, and then discuss how external stress induces quadrupoles inside a circular hole or inclusion.

Let us consider a perfectly conductive disk of radius $R$ in a uniform external electric field ($\mathbf{E}=E_0\boldsymbol{\mbf{\hat{x}}}$) in 2D. This electric field provides a driving force for mobile charges on the disk, which are  redistributed until the resulting tangential component of the total electric field at the circumference of the disk is zero. This means that the electric potential is constant on the circumference ($r=R$). Assuming that the electric potential is zero on the circumference of the disk and that the resultant electric field approaches the background field far away from the disk, we can solve the governing Eq.~(\ref{eq:GovElectrostatics}) with $\rho_e=0$ in polar coordinates to find that the electric potential is $U_\text{in}^\text{tot}(r,\varphi)=0$ inside the conductive disk ($r<R$) and that the electric potential $U_\text{out}^\text{tot}(r,\varphi)$ outside the conductive disk ($r>R$) is given by~\cite{Smythe}
\begin{equation}
  \label{eq:ElecInduction}
  \begin{split}
  U_\text{out}^\text{tot}(r,\varphi) &= -E_0 r\cos\varphi+E_0\frac{R^2}{r}\cos\varphi,
  \end{split}
\end{equation}
where the origin of the coordinate system is at the center of the conductive disk. The first term in the above Eq.~(\ref{eq:ElecInduction}) for the electric potential $U_\text{out}^\text{tot}(r,\varphi)$ outside the conductive disk is due to the external electric field and the second term can be interpreted as the electric potential of an induced electrostatic dipole at the center of the disk (see Eq.~(\ref{eq:ElecDipole}) and Fig.~\ref{fig:induction}a). This analysis can be generalized to a dielectric disk with dielectric constant $\epsilon_\text{in}$ that is embedded in a material with the dielectric constant $\epsilon_\text{out}$ in a uniform external electric field ($\mathbf{E}=E_0\boldsymbol{\mbf{\hat{x}}}$).
The electric potentials inside and outside the disk are then given by~\cite{Smythe}
 \begin{subequations}
  \label{eq:ElecInductiondielectric}
\begin{align}
  U_\text{in}^\text{tot}(r,\varphi) &= -E_0 r\cos\varphi+E_0 \frac{(\epsilon_\text{in}-\epsilon_\text{out})}{(\epsilon_\text{in}+\epsilon_\text{out})}r\cos\varphi,\\
  U_\text{out}^\text{tot}(r,\varphi) &= -E_0 r\cos\varphi+E_0\frac{(\epsilon_\text{in}-\epsilon_\text{out})}{(\epsilon_\text{in}+\epsilon_\text{out})}\frac{R^2}{r}\cos\varphi.
\end{align}
 \end{subequations}
The first terms in both $U_\text{in}^\text{tot}$ and $U_\text{out}^\text{tot}$ correspond to the external electric field, whereas the second terms can be interpreted as induced dipoles. The expression in Eq.~(\ref{eq:ElecInduction}) for the conductive disk is recovered in the limit $\epsilon_\text{in}/\epsilon_\text{out}~\rightarrow~\infty$. Note that the resulting electric field inside the dielectric disk is uniform $\mbf{E}^\text{tot}_\text{in}=-\bs{\nabla}U^\text{tot}_\text{in}=2 E_0 \epsilon_\text{out}/(\epsilon_\text{in}+\epsilon_\text{out}) \mbf{\hat x}$.

Similarly, external stress induces multipoles in elastic systems. For example, consider a circular hole of radius $R$ embedded in an infinite elastic matrix. Under external stress $\sigma^\textrm{ext}_{xx}=-\sigma_0$, the resultant Airy stress function is obtained by solving the governing  Eq.~(\ref{eq:GovElasticity}) with $\rho=0$ with the traction-free boundary condition ($\sigma_{rr}=\sigma_{r\varphi}=0$) at the circumference of the hole. The Airy stress function outside the hole ($r>R$) in polar coordinates is given by~\cite{Kamien}
\begin{equation}
  \label{eq:ElasInduction}
    \chi^\text{tot}_\text{out}(r,\varphi) = -\frac{\sigma_0 r^2}{4}\big(1-\cos (2\varphi)\big) \,+ \,{ \frac{\sigma_0 R^2}{2}\ln r }\,- \,{\frac{\sigma_0 R^2}{2}\cos (2\varphi)} \,+\,{\frac{\sigma_0 R^4}{4r^2}\cos(2\varphi)}.
\end{equation}
The above equation for the Airy stress function reveals that the external stress induces a non-topological monopole $p$ (Eq.~(\ref{eq:ElasMonopoleP})), and quadrupoles $\mbf{Q}^s$ and $\mbf{Q}^p$ (Eqs.~(\ref{eq:ElasQuadrupoleS},\ref{eq:ElasQuadrupoleP})) at the center of the hole (see Fig.~\ref{fig:induction}c). Note that unlike in electrostatics, dipoles $\mbf{d}^s$ are not induced in elasticity. This is because isolated disclinations (topological monopoles) and dislocations (topological dipoles) are formed by insertion or removal of material, which makes them topological defects~\cite{Chaikin}. On the other hand, elastic non-topological monopole $p$ and quadrupoles $\mbf{Q}^s$ and $\mbf{Q}^p$ can be obtained by local material rearrangement and can thus be induced by external loads~\cite{Moshe1,Moshe2}.

The above analysis can be generalized to the case with a circular inclusion of radius $R$ made from material with the Young's modulus $E_\textrm{in}$ and the Poisson's ratio $\nu_\textrm{in}$ that is embedded in an infinite elastic matrix made from material with the Young's modulus $E_\textrm{out}$ and the Poisson's ratio $\nu_\textrm{out}$. Under uniaxial compressive stress $\sigma^\textrm{ext}_{xx}=-\sigma_0$, the Airy stress function corresponding to the external stress is  $\chi_\text{ext}=-\sigma_0 y^2/2=-\sigma_0r^2\big(1-\cos(2\varphi)\big)/4$. Since the Airy stress function due to external stress contains both the axisymmetric and the $\cos(2\varphi)$ term, the Airy stress function due to induced multipoles should have the same angular dependence. Furthermore stresses should remain finite at the center of the inclusion ($r=0$) and also far away from the inclusion ($r \rightarrow \infty$).
The total Airy stress function $\chi^\text{tot}_\text{in}(r,\varphi)$ inside ($r<R$) and $\chi^\text{tot}_\text{out}(r,\varphi)$ outside ($r>R$) the inclusion can thus be written in the following form
\begin{subequations}
    \begin{align}
    \label{eq:InclusionAiryIn}
        \chi^\text{tot}_\text{in}(r,\varphi) = &-\frac{\sigma_0 r^2}{4}\big(1-\cos (2\varphi)\big)+c_0 r^2 +a_2 r^2 \cos(2\varphi)+c_2\frac{r^4}{R^2}\cos(2\varphi),\\
        \label{eq:InclusionAiryOut}
  \chi^\text{tot}_\text{out}(r,\varphi) = &-\frac{\sigma_0 r^2}{4}\big(1-\cos (2\varphi)\big)+A_0R^2\ln \left(\frac{r}{R}\right) +C_2 R^2\cos(2\varphi) + A_2 R^4 {r}^{-2}\cos(2\varphi).
    \end{align}
\end{subequations}
The last three terms in Eq.~(\ref{eq:InclusionAiryOut}) correspond to the induced non-topological monopole $p$ and quadrupoles $\mbf{Q}^s$ and $\mbf{Q}^p$ at the center of the inclusion, similar to induced multipoles at the center of the hole in Eq.~(\ref{eq:ElasInduction}). The last three terms in Eq.~(\ref{eq:InclusionAiryIn}) can also be interpreted as induced multipoles that are located far away from the inclusion. The unknown coefficients are determined from the boundary conditions, which require that tractions ($\sigma_{rr}$ and $\sigma_{r\varphi}$) and displacements ($u_r$ and $u_{\varphi}$) are continuous at the circumference of the inclusion ($r=R$). Stresses corresponding to the Airy stress function $\chi(r,\varphi)$ can be calculated as $\sigma_{rr}=r^{-1}(\partial \chi/\partial r)+r^{-2} (\partial^2\chi/\partial \varphi^2)$, $\sigma_{\varphi\varphi}=\partial^2\chi/\partial r^2$, and $\sigma_{r\varphi}=-\partial(r^{-1} \partial \chi/\partial \varphi)/\partial r$.  Table~\ref{table:MichellStressDisplacement}  summarizes the stresses corresponding to different terms in the Michell solution~\cite{Barber}.
\begingroup
\setlength{\tabcolsep}{2pt}
\begin{table}[t!]
{\centering
\caption{\label{table:MichellStressDisplacement}
Stresses $\sigma_{ij}$ and displacements $u_i$ corresponding to different terms for the Airy stress function $\chi$ in the Michell solution~\cite{Barber}. The value of Kolosov's constant for plane stress is $\kappa=(3-\nu)/(1+\nu)$ and for plane strain is  $\kappa=3-4\nu$. Here, $\mu$ is the shear modulus and $\nu$ is the Poisson's ratio.}
\footnotesize
\def\arraystretch{1.1}
\begin{tabular}{|c|c|c|c|c|}
\hline
$\chi$ & $\sigma_{rr}$ & $\sigma_{r\varphi}$ &  $\sigma_{\varphi\varphi}$ & $2\mu \begin{pmatrix}u_{r}\\u_{\varphi}\end{pmatrix}$ \\
\hline
$r^2$ & $2$ & $0$ & $2$ & $r \, \begin{pmatrix}\kappa-1\\
                                0\end{pmatrix}$ \\
$\ln r$ & $r^{-2}$ & $0$ & -$r^{-2}$ & $r^{-1}\, \begin{pmatrix}-1\\
                                0\end{pmatrix}$\\
$r^{n+2}\cos (n\varphi)$ & $-(n+1)(n-2)r^n\cos (n\varphi)$ & $n(n+1)r^n\sin (n\varphi)$ & $(n+1)(n+2)r^n\cos (n\varphi)$ & $r^{n+1}\, \begin{pmatrix}(\kappa-n-1)\cos (n\varphi) \\
                            (\kappa+n+1)\sin (n\varphi) \end{pmatrix}$\\
$r^{n+2}\sin (n\varphi)$ & $-(n+1)(n-2)r^n\sin (n\varphi)$ & $-n(n+1)r^n\cos (n\varphi)$ & $(n+1)(n+2)r^n\sin (n\varphi)$ & $r^{n+1}\, \begin{pmatrix}(\kappa-n-1)\sin (n\varphi) \\
                    -(\kappa+n+1)\cos (n\varphi)\end{pmatrix}$\\
$r^{-n+2}\cos (n\varphi)$ &$-(n+2)(n-1)r^{-n}\cos (n\varphi)$& $-n(n-1)r^{-n}\sin (n\varphi)$ & $(n-1)(n-2)r^{-n}\cos (n\varphi)$ &$r^{-n+1}\,\begin{pmatrix}(\kappa+n-1)\cos (n\varphi) \\
                        -(\kappa-n+1)\sin (n\varphi) \end{pmatrix}$\\
$r^{-n+2}\sin (n\varphi)$ & $-(n+2)(n-1)r^{-n}\sin (n\varphi)$ & $n(n-1)r^{-n}\cos (n\varphi)$ & $(n-1)(n-2)r^{-n}\sin (n\varphi)$ & $r^{-n+1}\, \begin{pmatrix}(\kappa+n-1)\sin (n\varphi) \\
                    (\kappa-n+1)\cos (n\varphi)\end{pmatrix}$\\
$r^{n}\cos (n\varphi)$ & $-n(n-1)r^{n-2}\cos (n\varphi)$ & $n(n-1)r^{n-2}\sin (n\varphi)$ & $n(n-1)r^{n-2}\cos (n\varphi)$ & $r^{n-1}\, \begin{pmatrix}-n \cos (n\varphi) \\
                        n \sin (n\varphi)\end{pmatrix}$\\
$r^{n}\sin (n\varphi)$ & $-n(n-1)r^{n-2}\sin (n\varphi)$ & $-n(n-1)r^{n-2}\cos (n\varphi)$ & $n(n-1)r^{n-2}\sin (n\varphi)$ &  $r^{n-1} \begin{pmatrix}-n \sin (n\varphi)\\
                        -n \cos (n\varphi)\end{pmatrix}$\\
$r^{-n}\cos (n\varphi)$ & $-n(n+1)r^{-n-2}\cos (n\varphi)$ & $-n(n+1)r^{-n-2}\sin (n\varphi)$ & $n(n+1)r^{-n-2}\cos (n\varphi)$ & $r^{-n-1}\, \begin{pmatrix}n \cos (n\varphi) \\
                        n \sin (n\varphi)\end{pmatrix}$\\
$r^{-n}\sin (n\varphi)$ & $-n(n+1)r^{-n-2}\sin (n\varphi)$ & $n(n+1)r^{-n-2}\cos (n\varphi)$ & $n(n+1)r^{-n-2}\sin (n\varphi)$ & $r^{-n-1}\,\begin{pmatrix}n \sin (n\varphi)\\-n \cos (n\varphi)\end{pmatrix}$\\
\hline
\end{tabular}
}
\end{table}
\endgroup
The boundary conditions for tractions ($\sigma_{rr}$ and $\sigma_{r\varphi}$) at the circumference of the inclusion are thus written as
\begin{equation}
\label{eq:BoundaryStress}
    \begin{split}
        -\frac{\sigma_0}{2} + 2 c_0 - \left(\frac{\sigma_0}{2}+2 a_2 \right) \cos (2\varphi) = & -\frac{\sigma_0}{2} + A_0 - \left(\frac{\sigma_0}{2}+4 C_2 + 6 A_2 \right) \cos (2\varphi),\\
         \left(\frac{\sigma_0}{2}+2 a_2 + 6 c_2\right) \sin (2\varphi) = &  \left(\frac{\sigma_0}{2}-2 C_2 - 6 A_2\right) \sin (2\varphi).
    \end{split}
\end{equation}
In order to obtain displacements, we first calculate the strains  $\varepsilon_{rr}=\left((\kappa+1)\sigma_{rr}-(3-\kappa) \sigma_{\varphi \varphi}\right)/(8\mu)$, $\varepsilon_{r\varphi}=\sigma_{r\varphi}/(2\mu)$ and $\varepsilon_{\varphi\varphi}=\left((\kappa+1)\sigma_{\varphi\varphi}-(3-\kappa) \sigma_{rr}\right)/(8\mu)$, 
where $\mu=E/[2(1+\nu)]$ is the shear modulus and we  introduced the Kolosov's constant $\kappa=(3-\nu)/(1+\nu)$ for  plane stress and  $\kappa=3-4\nu$ for  plane strain condition~\cite{Barber}. Displacements $u_r$ and $u_\varphi$ are then obtained by  integrating the strains. 
Table~\ref{table:MichellStressDisplacement}  summarizes the displacements corresponding to different terms in the Michell solution~\cite{Barber}. The boundary conditions for displacements ($u_r$ and $u_\varphi$) at the circumference of inclusion are thus written as
\begin{equation}
\label{eq:BoundaryDisplacement}
\begin{split}
&\left(-\frac{\sigma_0}{4} + c_0\right)\frac{R(\kappa_\textrm{in}-1)}{2 \mu_\textrm{in}} + \left(-\frac{\sigma_0}{2}-2 a_2 + c_2 (\kappa_\textrm{in}-3)\right) \frac{R \cos (2\varphi)}{2 \mu_\textrm{in}}=\\
&\hspace{3cm} -\frac{R}{{2 \mu_\textrm{out}}}\,\left(\frac{1}{4} \sigma_0 (\kappa_\textrm{out}-1) + A_0 \right) + \left(-\frac{\sigma_0}{2} + C_2 (\kappa_\textrm{out}+1) + 2 A_2\right)\frac{R \cos (2 \varphi)}{2 \mu_\textrm{out}},\\
&\left(\frac{\sigma_0}{2} + 2 a_2 +c_2 (\kappa_\textrm{in}+3)\right) \frac{R \sin (2 \varphi)}{2 \mu_\textrm{in}}=\left(\frac{\sigma_0}{2} - C_2 (\kappa_\textrm{out}-1)+2A_2\right) \frac{R \sin (2 \varphi)}{2 \mu_\textrm{out}}.
\end{split}
\end{equation}
The boundary conditions in Eqs.~(\ref{eq:BoundaryStress}) and (\ref{eq:BoundaryDisplacement}) have to be satisfied at every point ($\varphi$) on the circumference of the inclusion. Thus the coefficients of the Fourier components have to match on both sides of these equations, which allows us to rewrite the boundary conditions as a matrix equation
\begin{equation}
\begin{pmatrix}
1 & 0 & 0 & -2 & 0 & 0\\
0 & -6 & -4 & 0 & 2 & 0\\
0 & -6 & -2 & 0 & -2 & -6\\
-\frac{R}{2 \mu_\textrm{out}} & 0 & 0 & -\frac{ R(\kappa_\textrm{in}-1)}{2 \mu_\textrm{in}} & 0 & 0\\
0 & \frac{R}{\mu_\textrm{out}} & \frac{R(\kappa_\textrm{out}+1)}{2 \mu_\textrm{out}} & 0 & \frac{R}{\mu_\textrm{in}} & -\frac{ R(\kappa_\textrm{in}-3)}{2 \mu_\textrm{in}}\\
0 & \frac{R}{\mu_\textrm{out}} & -\frac{R(\kappa_\textrm{out}-1)}{2 \mu_\textrm{out}} & 0 & -\frac{R}{\mu_\textrm{in}} & -\frac{R (\kappa_\textrm{in}+3)}{2 \mu_\textrm{in}}
\end{pmatrix}
\begin{pmatrix}
A_0\\A_2\\C_2\\c_0\\a_2\\c_2
\end{pmatrix}
=
\begin{pmatrix}
0\\
0\\
0\\
\frac{\sigma_0 R}{8} \left(\frac{(\kappa_\textrm{out}-1)}{\mu_\textrm{out}} - \frac{(\kappa_\textrm{in}-1)}{\mu_\textrm{in}} \right)\\
\frac{\sigma_0 R}{4} \left(\frac{1}{\mu_\textrm{out}}-\frac{1}{\mu_\textrm{in}}\right)\\
\frac{\sigma_0 R}{4} \left(-\frac{1}{\mu_\textrm{out}}+\frac{1}{\mu_\textrm{in}}\right)
\end{pmatrix}.
\end{equation}
By solving the above set of equations we find that the Airy stress functions $\chi^\text{tot}_\text{in}(r,\varphi)$ inside ($r<R$) and $\chi^\text{tot}_\text{out}(r,\varphi)$ outside ($r>R$) the inclusion are given by
\begin{subequations}
  \label{eq:ElasInductioninclusion}
\begin{align}
  \chi^\text{tot}_\text{in}(r,\varphi) = &-\frac{\sigma_0 r^2}{4}\big(1-\cos (2\varphi)\big)+\frac{ (\mu_\text{out}(\kappa_\text{in}-1)-\mu_\text{in}(\kappa_\text{out}-1))}{4(\mu_\text{out}(\kappa_\text{in}-1)+2\mu_\text{in})}\,\sigma_0 r^2-\frac{(\mu_\text{out}-\mu_\text{in})}{4(\mu_\text{out}+ \mu_\text{in}\kappa_\text{out})}\,\sigma_0r^2\cos(2\varphi),\\
  \chi_\text{out}^\text{tot}(r,\varphi) = &-\frac{\sigma_0 r^2}{4}\big(1-\cos (2\varphi)\big)+\frac{(\mu_\text{out}(\kappa_\text{in}-1)-\mu_\text{in}(\kappa_\text{out}-1))}{2(\mu_\text{out}(\kappa_\text{in}-1)+2\mu_\text{in})}\,\sigma_0 R^2\ln r -\frac{(\mu_\text{out}-\mu_\text{in})}{2(\mu_\text{out}+ \mu_\text{in}\kappa_\text{out})}\,\sigma_0 R^2\cos(2\varphi)\nonumber \\
  &+\frac{(\mu_\text{out}-\mu_\text{in})}{4(\mu_\text{out}+\mu_\text{in} \kappa_\text{out})}\sigma_0 R^4 r^{-2}\cos(2\varphi).
\end{align}
\end{subequations}
In the above Eq.~(\ref{eq:ElasInductioninclusion}) for the Airy stress functions the last three terms can again be interpreted as induced non-topological monopole $p$ and quadrupoles $\mbf{Q}^s$ and $\mbf{Q}^p$. The expression in Eq.~(\ref{eq:ElasInduction}) for the hole is recovered in the limit $\mu_\textrm{in}\rightarrow 0$. Note that similar to the Eshelby inclusions in 3D~\cite{Eshelby}, the stress field inside the inclusion in 2D is uniform and is given by 
\begin{subequations}
\begin{align}
    \sigma^\textrm{in}_{xx}&=-\sigma_0\, \frac{\mu_\text{in}(1+\kappa_\text{out})(\mu_\text{out} \kappa_\text{in}+\mu_\text{in}(2+\kappa_\text{out}))}{2(\mu_\text{out}(\kappa_\text{in}-1)+2\mu_\text{in})(\mu_\text{out}+\mu_\text{in}\kappa_\text{out})},\\
\sigma^\textrm{in}_{yy}&=\sigma_0\, \frac{\mu_\text{in}(1+\kappa_\text{out})(\mu_\text{out}(\kappa_\text{in}-2)-\mu_\text{in}(\kappa_\text{out}-2))}{2(\mu_\text{out}(\kappa_\text{in}-1)+2\mu_\text{in})(\mu_\text{out}+\mu_\text{in}\kappa_\text{out})},\\
\sigma^\textrm{in}_{xy}&=0.
\end{align}
\label{eq:SingleInclusionStress}
\end{subequations}
By comparing the above analyses in elasticity and electrostatics, we conclude that holes and inclusions in elasticity are analogous to perfect conductors and dielectrics in electrostatics, respectively.

The problem of induction becomes much more involved when multiple dielectric objects are considered in electrostatics or multiple inclusions in elasticity. This is because dielectric objects and inclusions interact with each other via induced electric fields and stress fields, respectively. In the next Section, we describe how such interactions can be systematically taken into account in elasticity, which enabled us to calculate the magnitudes of induced multipoles in the presence of external load.

\section{Elastic multipole method}\label{sec:ElasticMultipoleMethod}
Building on the concepts described above, we have developed a method for calculating the linear deformation of circular inclusions and holes embedded in an infinite elastic matrix under external stress. External stress induces elastic multipoles at the centers of inclusions and holes, and their amplitudes are obtained from the boundary conditions between different materials (continuity of tractions and displacements). In the following Section \ref{sec:method}, we describe the method for the general case where circular inclusions can have different sizes and material properties (holes correspond to zero shear modulus). Note that our method applies to the deformation of cylindrical holes and inclusions embedded in thin plates (plane stress) as well as to cylindrical holes and inclusions embedded in an infinitely thick elastic matrix (plane strain) by appropriately setting the values of the Kolosov's constant. In Section \ref{sec:ExampleInfinite} we compare the results of our method to the finite element simulations and in Section \ref{sec:Experiments} they are compared to experiments.

\subsection{Method}\label{sec:method}
Let us consider a 2D infinite elastic matrix with the Young's modulus $E_0$ and the Poisson's ratio $\nu_0$. Embedded in the matrix are $N$ circular inclusions with radii $R_i$ centered at positions $\mbf{x}_i=(x_i,y_i)$ with Young's moduli $E_i$ and Poisson's ratios $\nu_i$, where $i\in\{1,\ldots, N\}$. Holes are described with zero Young's modulus ($E_i=0$). External stress, represented with the Airy stress function
\begin{equation}
    \chi_\textrm{ext}(x,y)=\frac{1}{2}\sigma_{xx}^\textrm{ext} y^2 + \frac{1}{2}\sigma_{yy}^\textrm{ext} x^2 - \sigma_{xy}^\textrm{ext} xy,
    \label{eq:ChiExt}
\end{equation}
induces non-topological monopoles ($p$), non-topological dipoles ($\mbf{d}^p$), quadrupoles ($\mbf{Q}^s$, $\mbf{Q}^p$), and higher-order multipoles ($\mbf{Q}^s_n$, $\mbf{Q}^p_n$) at the centers of
inclusions, as was discussed in Section~\ref{sec:induction}. Thus the Airy stress function outside the $i^{\text{th}}$ inclusion due to the induced multipoles can be expanded as
\begin{equation}
  \label{eq:AiryInducedOut}
  \begin{split}
  \chi_{\text{out}}\big(r_i,\varphi_i|\mbf{a}_{i,\text{out}}\big) = &A_{i,0} R_i^2 \ln \left(\frac{r_i}{R_i}\right) + \sum_{n=1}^\infty R_i^2 \left[ A_{i,n} \left(\frac{r_i}{R_i}\right)^{-n}\cos(n\varphi_i) +B_{i,n}
  \left(\frac{r_i}{R_i}\right)^{-n}\sin(n \varphi_i)\right]\\
  &\quad \quad \quad \quad \quad\quad\quad  +\sum_{n=2}^{\infty} R_i^2 \left[ C_{i,n} \left(\frac{r_i}{R_i}\right)^{-n+2}\cos (n
  \varphi_i) + D_{i,n}
  \left(\frac{r_i}{R_i}\right)^{-n+2}\sin (n \varphi_i)\right],
  \end{split}
\end{equation}
where the origin of polar coordinates $(r_i = \sqrt{(x-x_i)^2+(y-y_i)^2},\varphi_i = \arctan[(y-y_i)/(x-x_i)])$ is at the center $\mbf{x}_i$ of the $i^\text{th}$ inclusion and $\mbf{a}_{i,\text{out}}=\{A_{i,0}, A_{i,1},\dots,B_{i,1}, B_{i,2},\dots,C_{i,2}, C_{i,3},\dots, D_{i,2}, D_{i,3},\dots\}$ is the set of amplitudes of induced multipoles. The total Airy stress function outside all inclusions can then be written as
\begin{equation}
\label{eq:AiryOut}
\chi^{\text{tot}}_{\text{out}}\big(x,y|\mbf{a}_{\text{out}}\big)=\chi_{\text{ext}}(x,y)+\sum_{i=1}^N \chi_{\text{out}}\big(r_i(x,y),\varphi_i(x,y)|\mbf{a}_{i,\text{out}}\big),
\end{equation}
where the first term is due to external stress and the summation describes contributions due to induced multipoles at the centers of inclusions.  The set of amplitudes of induced multipoles for all inclusions is defined as $\mbf{a}_{\text{out}}=\{\mbf{a}_{\text{1,out}}, \cdots, \mbf{a}_{N,\text{out}}\}$.

Similarly, we expand the induced Airy stress function inside the $i^\text{th}$ inclusion as
\begin{equation}
  \label{eq:AiryInducedIn}
  \begin{split}
  \chi_{\text{in}}\big(r_i,\varphi_i|\mbf{a}_{i,\text{in}}\big)&= \phantom{+c_{i,0} r_i^2+}    \sum_{n=2}^{\infty} R_i^2 \left[a_{i,n} \left(\frac{r_i}{R_i}\right)^n\cos(n\varphi_i)+ b_{i,n}
  \left(\frac{r_i}{R_i}\right)^n\sin(n\varphi_i)\right]\\
  &\phantom{=} +c_{i,0} r_i^2 +\sum_{n=1}^{\infty} R_i^2 \left[c_{i,n} \left(\frac{r_i}{R_i}\right)^{n+2}\cos
  (n\varphi_i) + d_{i,n}
  \left(\frac{r_i}{R_i}\right)^{n+2}\sin (n\varphi_i) \right],
  \end{split}
\end{equation}
where we kept only the terms that generate finite stresses at the center of inclusion and omitted constant and linear terms $\{1, r_i \cos\varphi_i, r_i \sin \varphi_i\}$ that correspond to zero stresses. The set of amplitudes of induced multipoles is represented as $\mbf{a}_{i,\text{in}}=\{a_{i,2}, a_{i,3},\dots,b_{i,2}, b_{i,3},\dots, c_{i,0}, c_{i,1},\dots, d_{i,1}, d_{i,2},\dots\}$. The total Airy stress function inside the $i^\text{th}$ inclusion is thus
\begin{equation}
    \chi^{\text{tot}}_{\text{in}}\big(x,y|\mbf{a}_{i,\text{in}}\big)=\chi_{\text{ext}}(x,y)+\chi_{\text{in}}\big(r_i(x,y),\varphi_i(x,y)|\mbf{a}_{i,\text{in}}\big),
    \label{eq:AiryIn}    
\end{equation}
where the first term is due to external stress and the second term is due to induced multipoles.

\begin{figure}[t!]
  \centering
  \includegraphics[scale=1]{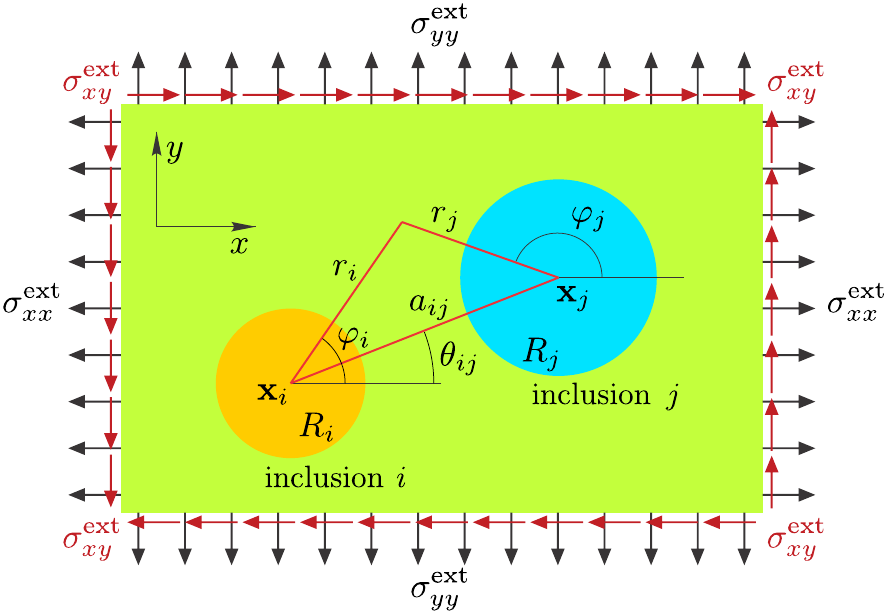}
  \caption{Illustration of external load ($\sigma_{xx}^\text{ext}$, $\sigma_{yy}^\text{ext}$, $\sigma_{xy}^\text{ext}$) and  polar coordinates ($r_i,\varphi_i$) and ($r_j,\varphi_j$) relative to the centers $\mbf{x}_i$ of the $i^{\text{th}}$ inclusion (orange disk) with radius $R_i$ and $\mbf{x}_j$ of the $j^{\text{th}}$ inclusion (blue disk) with radius $R_j$, respectively. Here, $a_{ij}$ is the separation distance between the $i^{\text{th}}$ and $j^{\text{th}}$ inclusion and $\theta_{ij}$ is the angle between the line joining their centers and the $x$-axis.}
  \label{Fig:ElasMultIllus}
\end{figure}

The amplitudes of induced multipoles $\mbf{a}_{i,\text{out}}$ and $\mbf{a}_{i,\text{in}}$ are obtained by satisfying the boundary conditions that tractions and displacements are continuous across the circumference of each inclusion
\begin{subequations}
  \begin{align}
    \sigma_{\text{in},rr}^{\text{tot}}\big(r_i=R_i,\varphi_i|\mbf{a}_{i,\text{in}}\big)&=\sigma_{\text{out},rr}^{\text{tot}}\big(r_i=R_i,\varphi_i|\mbf{a}_{\text{out}}\big), \label{eq:BC:sigmaRR}\\
    \sigma_{\text{in},r\varphi}^{\text{tot}}\big(r_i=R_i,\varphi_i|\mbf{a}_{i,\text{in}}\big)&=\sigma_{\text{out},r\varphi}^{\text{tot}}\big(r_i=R_i,\varphi_i|\mbf{a}_{\text{out}}\big), \label{eq:BC:sigmaRF}\\
    u_{\text{in},r}^{\text{tot}}\big(r_i=R_i,\varphi_i|\mbf{a}_{i,\text{in}}\big)&=u_{\text{out},r}^{\text{tot}}\big(r_i=R_i,\varphi_i|\mbf{a}_{\text{out}}\big), \label{eq:BC:uR}\\
    u_{\text{in},\varphi}^{\text{tot}}\big(r_i=R_i,\varphi_i|\mbf{a}_{i,\text{in}}\big)&=u_{\text{out},\varphi}^{\text{tot}}\big(r_i=R_i,\varphi_i|\mbf{a}_{\text{out}}\big),\label{eq:BC:uF}
   \end{align}
   \label{eq:BC}%
\end{subequations}
where stresses and displacements are obtained from the total Airy stress functions $\chi^{\text{tot}}_{\text{in}}\big(x,y|\mbf{a}_{i,\text{in}}\big)$ inside the $i^\text{th}$ inclusion (see Eq.~(\ref{eq:AiryIn})) and $\chi^{\text{tot}}_{\text{out}}\big(x,y|\mbf{a}_{\text{out}}\big)$ outside all inclusions (see Eq.~(\ref{eq:AiryOut})). In the boundary conditions for the $i^\text{th}$ inclusion in the above Eq.~(\ref{eq:BC}), we can easily take into account contributions due to the induced multipoles $\mbf{a}_{i,\text{in}}$ and $\mbf{a}_{i,\text{out}}$ in this inclusion and due to external stresses $\sigma_{xx}^\textrm{ext}$, $\sigma_{yy}^\textrm{ext}$, and $\sigma_{xy}^\textrm{ext}$ after rewriting the corresponding Airy stress function $\chi_\textrm{ext}(x,y)$ in Eq.~(\ref{eq:ChiExt}) in polar coordinates centered at the  $i^\text{th}$ inclusion as
\begin{equation}
\label{eq:ChiExt_polar}
\chi_\textrm{ext}(r_i,\varphi_i)=\frac{1}{4}(\sigma_{xx}^\textrm{ext}+\sigma_{yy}^\textrm{ext})r_i^2  - \frac{1}{4}(\sigma_{xx}^\textrm{ext}-\sigma_{yy}^\textrm{ext}) r_i^2 \cos(2 \varphi_i) - \frac{1}{2}\sigma_{xy}^\textrm{ext} r_i^2 \sin(2 \varphi_i).
\end{equation}
Contributions to stresses and displacements in the boundary conditions for the $i^\text{th}$ inclusion in Eq.~(\ref{eq:BC}) due to the Airy stress functions $\chi_{\text{in}}\big(r_i,\varphi_i|\mbf{a}_{i,\text{in}}\big)$, $\chi_{\text{out}}\big(r_i,\varphi_i|\mbf{a}_{i,\text{out}}\big)$, and $\chi_\textrm{ext}(r_i,\varphi_i)$ can be taken into account with the help of Table~\ref{table:MichellStressDisplacement}. However, it is not straightforward to consider the contributions due to the induced multipoles $\mbf{a}_{j,\text{out}}$ for other inclusions ($j\ne i$), because the corresponding Airy stress functions $\chi_{\text{out}}\big(r_j,\varphi_j|\mbf{a}_{j,\text{out}}\big)$ in Eq.~(\ref{eq:AiryInducedOut}) are written in the polar coordinates centered at $\mbf{x}_j$. The polar coordinates $(r_j,\varphi_j)$ centered at the $j^\text{th}$ inclusion can be expressed in terms of polar coordinates $(r_i,\varphi_i)$ centered at the $i^\text{th}$ inclusion as
\begin{equation}
\begin{split}
r_j(r_i,\varphi_i)&=\sqrt{r_i^2+a_{ij}^2-2 r_i a_{ij} \cos(\varphi_i - \theta_{ij})},\\
\varphi_j(r_i,\varphi_i)& =\pi +\theta_{ij} - \arctan\left[\frac{r_i \sin(\varphi_i  - \theta_{ij})}{\big(a_{ij}-r_i \cos(\varphi_i  - \theta_{ij})\big)}\right],
\end{split}
\end{equation}
where $a_{ij}=\sqrt{(x_i-x_j)^2+(y_i-y_j)^2}$ is the distance between the centers of the $i^{\text{th}}$ and $j^{\text{th}}$ inclusion and $\theta_{ij}=\arctan [(y_j-y_i)/(x_j-x_i)]$ is the angle between the line joining the centers of inclusions and the
$x$-axis, as shown in Fig.~\ref{Fig:ElasMultIllus}. 
The Airy stress function due to the induced multipoles centered at the $j^{\text{th}}$ inclusion can be expanded in Taylor series around the center of the $i^{\text{th}}$ inclusion as~\cite{Green}
\begin{equation}
\begin{split}
\chi_\text{out}\big(r_j(r_i,\varphi_i), \varphi_j(r_i,\varphi_i)|\mbf{a}_\textrm{j,out}\big)=&\quad \sum_{n=2}^\infty R_j^2 \frac{r_i^n}{a_{ij}^n} \Big[\cos(n \varphi_i) f^n_c\left(R_j/a_{ij},\theta_{ij}|\mbf{a}_{j,\textrm{out}}\right)+\sin(n \varphi_i) f^n_s\left(R_j/a_{ij},\theta_{ij}|\mbf{a}_{j,\textrm{out}}\right)\Big]\\
&
+\sum_{n=0}^\infty R_j^2\frac{r_i^{n+2}}{a_{ij}^{n+2}} \Big[\cos(n \varphi_i) g^n_c\left(R_j/a_{ij},\theta_{ij}|\mbf{a}_{j,\textrm{out}}\right)+\sin(n \varphi_i) g^n_s\left(R_j/a_{ij},\theta_{ij}|\mbf{a}_{j,\textrm{out}}\right)\Big],
\end{split}
\label{eq:AiryOutTaylorExpansion}
\end{equation}
where we omitted constant and linear terms $\{1, r_i \cos\varphi_i, r_i \sin \varphi_i\}$ that correspond to zero stresses and we introduced functions
\begin{subequations}
    \begin{align}
f^n_c\left(R_j/a_{ij},\theta_{ij}|\mbf{a}_{j,\textrm{out}}\right)&=\sum_{m=0}^\infty  \left(\frac{R_j^m}{a_{ij}^m} \Big[A_{j,m} \mathcal{A}_n^m(\theta_{ij})+B_{j,m} \mathcal{B}_n^m(\theta_{ij})\Big]+\frac{R_j^{m-2}}{a_{ij}^{m-2}}\Big[C_{j,m} \mathcal{C}_n^m(\theta_{ij})+D_{j,m} \mathcal{D}_n^m(\theta_{ij})\Big]\right),\\
f^n_s\left(R_j/a_{ij},\theta_{ij}|\mbf{a}_{j,\textrm{out}}\right)&=\sum_{m=0}^\infty  \left(\frac{R_j^m}{a_{ij}^m} \Big[A_{j,m} \mathcal{B}_n^m(\theta_{ij})-B_{j,m} \mathcal{A}_n^m(\theta_{ij})\Big]+\frac{R_j^{m-2}}{a_{ij}^{m-2}}\Big[C_{j,m} \mathcal{D}_n^m(\theta_{ij})-D_{j,m} \mathcal{C}_n^m(\theta_{ij})\Big]\right),\\
g^n_c\left(R_j/a_{ij},\theta_{ij}|\mbf{a}_{j,\textrm{out}}\right)&=\sum_{m=2}^\infty\frac{R_j^{m-2}}{a_{ij}^{m-2}}\Big[C_{j,m} \mathcal{E}_n^m(\theta_{ij})+D_{j,m} \mathcal{F}_n^m(\theta_{ij})\Big],\\
g^n_s\left(R_j/a_{ij},\theta_{ij}|\mbf{a}_{j,\textrm{out}}\right)&=\sum_{m=2}^\infty\frac{R_j^{m-2}}{a_{ij}^{m-2}}\Big[C_{j,m} \mathcal{F}_n^m(\theta_{ij})-D_{j,m} \mathcal{E}_n^m(\theta_{ij})\Big].
    \end{align}
    \label{eq:TaylorExpansion}%
\end{subequations}
In the above Eq.~(\ref{eq:TaylorExpansion}), we set $B_{j,0}=C_{j,0}=D_{j,0}=C_{j,1}=D_{j,1}=0$ and  introduced coefficients $\mathcal{A}_n^m(\theta_{ij})$, $\mathcal{B}_n^m(\theta_{ij})$, $\mathcal{C}_n^m(\theta_{ij})$, $\mathcal{D}_n^m(\theta_{ij})$, $\mathcal{E}_n^m(\theta_{ij})$, and $\mathcal{F}_n^m(\theta_{ij})$ that are summarized in Table~\ref{tab:CoefficientsFG}.
\begin{table}[!t]
{
\centering
\caption{Coefficients for the expansion of the Airy stress function $\chi_\text{out}\big(r_j(r_i,\varphi_i), \varphi_j(r_i,\varphi_i)|\mbf{a}_\textrm{j,out}\big)$ in Eqs.~(\ref{eq:AiryOutTaylorExpansion}-\ref{eq:TaylorExpansion}).}
\label{tab:CoefficientsFG}
\def\arraystretch{2}
\begin{tabular}{|l|l|l|}
\hline
$n \ge 2$&  $\mathcal{A}_n^0(\theta_{ij})=-\frac{1}{n} \cos(n\theta_{ij})$ & $\mathcal{B}_n^0(\theta_{ij})=-\frac{1}{n} \sin(n\theta_{ij})$\\ 
\hline
$n \ge 2, m \ge 1$ & $\mathcal{A}_n^m(\theta_{ij})=(-1)^m {{m+n-1}\choose{n}} \cos\big((m+n)\theta_{ij}\big)$ &  $\mathcal{B}_n^m(\theta_{ij})=(-1)^m {{m+n-1}\choose{n}} \sin\big((m+n)\theta_{ij}\big)$ \\
\hline
$n \ge 0, m \ge 2$ & $\mathcal{C}_n^m(\theta_{ij})=(-1)^m {{m+n-2}\choose{n}} \cos\big((m+n)\theta_{ij}\big)$ & $\mathcal{D}_n^m(\theta_{ij})=(-1)^m {{m+n-2}\choose{n}} \sin\big((m+n)\theta_{ij}\big)$\\
\hline
$n \ge 0, m \ge 2$ &$\mathcal{E}_n^m(\theta_{ij})=(-1)^{m-1} {{m+n-1}\choose{n+1}} \cos\big((m+n)\theta_{ij}\big)$ & $\mathcal{F}_n^m(\theta_{ij})=(-1)^{m-1} {{m+n-1}\choose{n+1}} \sin\big((m+n)\theta_{ij}\big)$\\
\hline
\end{tabular}
}
\end{table}

Next, we calculate stresses and displacements at the circumference of the $i^\text{th}$ inclusion by using expressions for the Airy stress functions \externalStress{$\chi_\textrm{ext}(r_i,\varphi_i)$} due to external stresses in Eq.~(\ref{eq:ChiExt_polar}), \inclusionI{$\chi_{\text{in}}\big(r_i,\varphi_i|\mbf{a}_{i,\text{in}}\big)$} and \inclusionI{$\chi_{\text{out}}\big(r_i,\varphi_i|\mbf{a}_{i,\text{out}}\big)$} in Eqs.~(\ref{eq:AiryInducedIn}) and (\ref{eq:AiryInducedOut}) due to induced multipoles for the $i^\text{th}$ inclusion, and \inclusionJ{$\chi_{\text{out}}\big(r_j,\varphi_j|\mbf{a}_{j,\text{out}}\big)$} in Eq.~(\ref{eq:AiryOutTaylorExpansion}) due to the induced multipoles for the $j^\text{th}$ inclusion ($j\ne i$) . With the help of Table~\ref{table:MichellStressDisplacement}, which shows how to convert each term of the Airy stress function to stresses and displacements, we obtain
\begin{subequations}
\begin{align}
\sigma_{\text{in},rr}^{\text{tot}}\big(r_i=R_i,\varphi_i|\mbf{a}_{i,\text{in}}\big)&=\externalStress{\frac{1}{2}(\sigma_{xx}^\textrm{ext} + \sigma_{yy}^\textrm{ext}) + \frac{1}{2}(\sigma_{xx}^\textrm{ext} - \sigma_{yy}^\textrm{ext}) \cos(2 \varphi_i) + \sigma_{xy}^\textrm{ext} \sin (2 \varphi_i)}+ \inclusionI{2 c_{i,0}} \nonumber\\
      &\hspace{-2.5cm}\inclusionI{-\sum_{n=1}^\infty \Big[n (n-1) \big(a_{i,n} \cos (n \varphi_i) + b_{i,n} \sin (n \varphi_i) \big)+(n+1)(n-2)\big(c_{i,n} \cos (n \varphi_i) + d_{i,n} \sin (n \varphi_i) \big) \Big]},\\
\sigma_{\text{out},rr}^{\text{tot}}\big(r_i=R_i,\varphi_i|\mbf{a}_{\text{out}}\big)&=  \externalStress{\frac{1}{2}(\sigma_{xx}^\textrm{ext} + \sigma_{yy}^\textrm{ext}) + \frac{1}{2}(\sigma_{xx}^\textrm{ext} - \sigma_{yy}^\textrm{ext}) \cos(2 \varphi_i) + \sigma_{xy}^\textrm{ext} \sin (2 \varphi_i)} + \inclusionI{A_{i,0}}\nonumber \\
    &\hspace{-2.5cm}\inclusionI{-\sum_{n=1}^\infty \Big[n (n+1) \big(A_{i,n} \cos (n \varphi_i) + B_{i,n} \sin (n \varphi_i) \big)+(n+2)(n-1)\big(C_{i,n} \cos (n \varphi_i) + D_{i,n} \sin (n \varphi_i) \big) \Big]}\nonumber\\
    &\hspace{-2.5cm}\inclusionJ{-\sum_{j\ne i}\sum_{n=2}^\infty \frac{R_j^2 R_i^{n-2} }{a_{ij}^n} n (n-1)  \Big[\cos(n \varphi_i) f^n_c\big(R_j/a_{ij},\theta_{ij}|\mbf{a}_{j,\textrm{out}}\big)+\sin(n \varphi_i) f^n_s\big(R_j/a_{ij},\theta_{ij}|\mbf{a}_{j,\textrm{out}}\big)\Big]}\nonumber \\
    &\hspace{-2.5cm}\inclusionJ{-\sum_{j\ne i}\sum_{n=0}^\infty \frac{R_j^2 R_i^{n}}{a_{ij}^{n+2}}  (n+1) (n-2) \Big[\cos(n \varphi_i) g^n_c\big(R_j/a_{ij},\theta_{ij}|\mbf{a}_{j,\textrm{out}}\big)+\sin(n \varphi_i) g^n_s\big(R_j/a_{ij},\theta_{ij}|\mbf{a}_{j,\textrm{out}}\big)\Big]},\\
\sigma_{\text{in},r\varphi}^{\text{tot}}\big(r_i=R_i,\varphi_i|\mbf{a}_{i,\text{in}}\big)&= \externalStress{-\frac{1}{2}(\sigma_{xx}^\textrm{ext} - \sigma_{yy}^\textrm{ext}) \sin(2 \varphi_i) + \sigma_{xy}^\textrm{ext} \cos (2 \varphi_i)} \nonumber\\
      &\hspace{-2.5cm}\inclusionI{+\sum_{n=1}^\infty \Big[n (n-1) \big(a_{i,n} \sin (n \varphi_i) - b_{i,n} \cos (n \varphi_i) \big)+n(n+1)\big(c_{i,n} \sin (n \varphi_i) - d_{i,n} \cos (n \varphi_i) \big) \Big]},\\
\sigma_{\text{out},r\varphi}^{\text{tot}}\big(r_i=R_i,\varphi_i|\mbf{a}_{\text{out}}\big)&= \externalStress{ -\frac{1}{2}(\sigma_{xx}^\textrm{ext} - \sigma_{yy}^\textrm{ext}) \sin(2 \varphi_i) + \sigma_{xy}^\textrm{ext} \cos (2 \varphi_i)}\nonumber \\
    &\hspace{-2.5cm}\inclusionI{-\sum_{n=1}^\infty \Big[n (n+1) \big(A_{i,n} \sin (n \varphi_i) - B_{i,n} \cos (n \varphi_i) \big)+n(n-1)\big(C_{i,n} \sin (n \varphi_i) - D_{i,n} \cos (n \varphi_i) \big) \Big]}\nonumber\\
    &\hspace{-2.5cm}\inclusionJ{+\sum_{j\ne i}\sum_{n=2}^\infty \frac{R_j^2 R_i^{n-2} }{a_{ij}^n} n (n-1)  \Big[\sin(n \varphi_i) f^n_c\big(R_j/a_{ij},\theta_{ij}|\mbf{a}_{j,\textrm{out}}\big)-\cos(n \varphi_i) f^n_s\big(R_j/a_{ij},\theta_{ij}|\mbf{a}_{j,\textrm{out}}\big)\Big]}\nonumber \\
    &\hspace{-2.5cm}\inclusionJ{+\sum_{j\ne i}\sum_{n=0}^\infty \frac{R_j^2 R_i^{n}}{a_{ij}^{n+2}}  n(n+1) \Big[\sin(n \varphi_i) g^n_c\big(R_j/a_{ij},\theta_{ij}|\mbf{a}_{j,\textrm{out}}\big)-\cos(n \varphi_i) g^n_s\big(R_j/a_{ij},\theta_{ij}|\mbf{a}_{j,\textrm{out}}\big)\Big]},\\
\frac{2\mu_i}{R_i}\,  u_{\text{in},r}^{\text{tot}}\big(r_i=R_i,\varphi_i|\mbf{a}_{i,\text{in}}\big)&= \externalStress{\frac{1}{4}(\sigma_{xx}^\textrm{ext} + \sigma_{yy}^\textrm{ext})(\kappa_i-1) + \frac{1}{2}(\sigma_{xx}^\textrm{ext} - \sigma_{yy}^\textrm{ext}) \cos(2 \varphi_i) + \sigma_{xy}^\textrm{ext}\sin (2 \varphi_i)} +\inclusionI{c_{i,0} (\kappa_i-1)} \nonumber\\
      &\hspace{-2.5cm}\inclusionI{-\sum_{n=1}^\infty \Big[n \big(a_{i,n} \cos (n \varphi_i) + b_{i,n} \sin (n \varphi_i) \big)+(n+1-\kappa_i)\big(c_{i,n} \cos (n \varphi_i) + d_{i,n} \sin (n \varphi_i) \big) \Big]},\\
\frac{2\mu_0}{R_i}\, u_{\text{out},r}^{\text{tot}}\big(r_i=R_i,\varphi_i|\mbf{a}_{\text{out}}\big)&= \externalStress{ \frac{1}{4}(\sigma_{xx}^\textrm{ext} + \sigma_{yy}^\textrm{ext})(\kappa_0-1) + \frac{1}{2}(\sigma_{xx}^\textrm{ext} - \sigma_{yy}^\textrm{ext}) \cos(2 \varphi_i) + \sigma_{xy}^\textrm{ext} \sin (2 \varphi_i)} \inclusionI{- A_{i,0}}\nonumber \\
    &\hspace{-2.5cm}\inclusionI{+\sum_{n=1}^\infty \Big[n \big(A_{i,n} \cos (n \varphi_i) + B_{i,n} \sin (n \varphi_i) \big)+(\kappa_0+n-1)\big(C_{i,n} \cos (n \varphi_i) + D_{i,n} \sin (n \varphi_i) \big) \Big]}\nonumber\\
    &\hspace{-2.5cm}\inclusionJ{-\sum_{j\ne i}\sum_{n=2}^\infty \frac{R_j^2 R_i^{n-2} }{a_{ij}^n} n  \Big[\cos(n \varphi_i) f^n_c\big(R_j/a_{ij},\theta_{ij}|\mbf{a}_{j,\textrm{out}}\big)+\sin(n \varphi_i) f^n_s\big(R_j/a_{ij},\theta_{ij}|\mbf{a}_{j,\textrm{out}}\big)\Big]}\nonumber \\
    &\hspace{-2.5cm}\inclusionJ{+\sum_{j\ne i}\sum_{n=0}^\infty \frac{R_j^2 R_i^{n}}{a_{ij}^{n+2}}  (\kappa_0-n-1) \Big[\cos(n \varphi_i) g^n_c\big(R_j/a_{ij},\theta_{ij}|\mbf{a}_{j,\textrm{out}}\big)+\sin(n \varphi_i) g^n_s\big(R_j/a_{ij},\theta_{ij}|\mbf{a}_{j,\textrm{out}}\big)\Big]},\\
\frac{2\mu_i}{R_i}\, u_{\text{in},\varphi}^{\text{tot}}\big(r_i=R_i,\varphi_i|\mbf{a}_{i,\text{in}}\big)&= \externalStress{-\frac{1}{2}(\sigma_{xx}^\textrm{ext} - \sigma_{yy}^\textrm{ext}) \sin(2 \varphi_i) + \sigma_{xy}^\textrm{ext} \cos (2 \varphi_i)} \nonumber\\
      &\hspace{-2.5cm}\inclusionI{+\sum_{n=1}^\infty \Big[n  \big(a_{i,n} \sin (n \varphi_i) - b_{i,n} \cos (n \varphi_i) \big)+(\kappa_i+n+1)\big(c_{i,n} \sin (n \varphi_i) - d_{i,n} \cos (n \varphi_i) \big) \Big]},\\
\frac{2\mu_0}{R_i}\, u_{\text{out},\varphi}^{\text{tot}}\big(r_i=R_i,\varphi_i|\mbf{a}_{\text{out}}\big)&= \externalStress{ -\frac{1}{2}(\sigma_{xx}^\textrm{ext} - \sigma_{yy}^\textrm{ext}) \sin(2 \varphi_i) + \sigma_{xy}^\textrm{ext} \cos (2 \varphi_i)}\nonumber \\
    &\hspace{-2.5cm}\inclusionI{+\sum_{n=1}^\infty \Big[n  \big(A_{i,n} \sin (n \varphi_i) - B_{i,n} \cos (n \varphi_i) \big)-(\kappa_0-n+1)\big(C_{i,n} \sin (n \varphi_i) - D_{i,n} \cos (n \varphi_i) \big) \Big]}\nonumber\\
    &\hspace{-2.5cm}\inclusionJ{+\sum_{j\ne i}\sum_{n=2}^\infty \frac{R_j^2 R_i^{n-2} }{a_{ij}^n} n \Big[\sin(n \varphi_i) f^n_c\big(R_j/a_{ij},\theta_{ij}|\mbf{a}_{j,\textrm{out}}\big)-\cos(n \varphi_i) f^n_s\big(R_j/a_{ij},\theta_{ij}|\mbf{a}_{j,\textrm{out}}\big)\Big]}\nonumber \\
    &\hspace{-2.5cm}\inclusionJ{+\sum_{j\ne i}\sum_{n=0}^\infty \frac{R_j^2 R_i^{n}}{a_{ij}^{n+2}}  (\kappa_0+n+1) \Big[\sin(n \varphi_i) g^n_c\big(R_j/a_{ij},\theta_{ij}|\mbf{a}_{j,\textrm{out}}\big)-\cos(n \varphi_i) g^n_s\big(R_j/a_{ij},\theta_{ij}|\mbf{a}_{j,\textrm{out}}\big)\Big]}.
\end{align}
\label{eq:BoundaryTractionsDisplacements}%
\end{subequations}
Colors in the above equations correspond to the Airy stress functions \externalStress{$\chi_\textrm{ext}(r_i,\varphi_i)$}, \inclusionI{$\chi_{\text{in}}\big(r_i,\varphi_i|\mbf{a}_{i,\text{in}}\big)$}, \inclusionI{$\chi_{\text{out}}\big(r_i,\varphi_i|\mbf{a}_{i,\text{out}}\big)$}, and  \inclusionJ{$\chi_{\text{out}}\big(r_j,\varphi_j|\mbf{a}_{j,\text{out}}\big)$}. We introduced the shear modulus $\mu_i=E_i/[2(1+\nu_i)]$ and  the Kolosov's constant $\kappa_i$ for the $i^\text{th}$ inclusion, where the value of Kolosov's constant is $\kappa_i=(3-\nu_i)/(1+\nu_i)$ for plane stress and $\kappa_i=3-4\nu_i$ for plane strain conditions~\cite{Barber}. Similarly, we define the shear modulus $\mu_0=E_0/[2(1+\nu_0)]$ and the Kolosov's constant $\kappa_0$ for the elastic matrix.

\begin{figure}[!b]
  \includegraphics[scale=1]{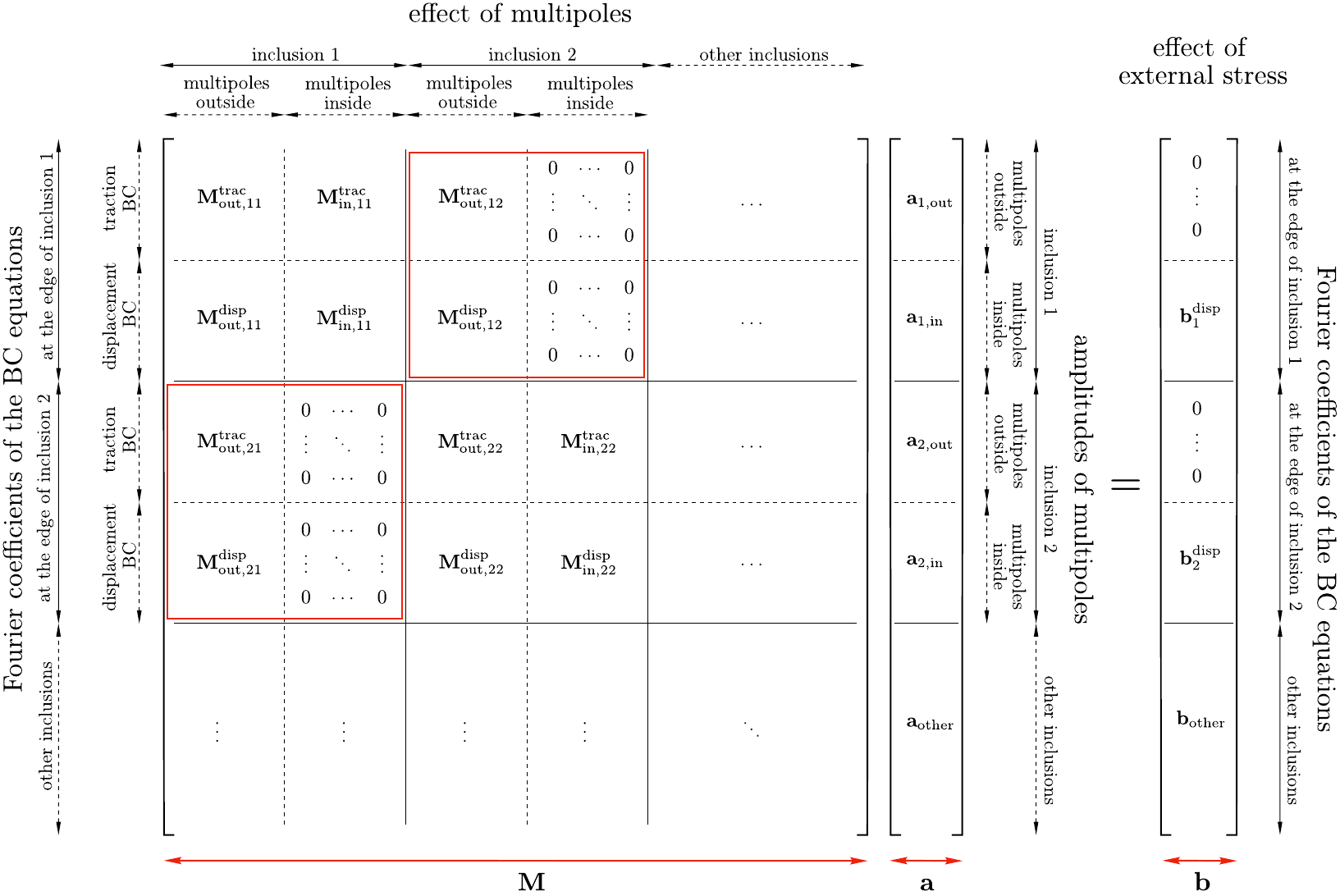}
  \caption{Structure of the system of Equations~(\ref{eq:matrix}) for the amplitudes $\mbf{a}_{i,\textrm{out}}$ and $ \mbf{a}_{i,\textrm{in}}$ of the induced multipoles for inclusions $i\in\{1,\ldots, N\}$. The matrix $\mathbf{M}$ is  divided into $4 N^2$ blocks, where the blocks $\mbf{M}_{\text{out},ij}^\text{trac}$ and $\mbf{M}_{\text{in},ij}^\text{trac}$ correspond to the boundary conditions for tractions around the circumference of the $i^\text{th}$ inclusion in Eq.~(\ref{eq:BC:sigmaRR}, \ref{eq:BC:sigmaRF}), and the blocks $\mbf{M}_{\text{out},ij}^\text{disp}$ and  $\mbf{M}_{\text{in},ij}^\text{disp}$ correspond to the boundary conditions for displacements around the circumference of the $i^\text{th}$ inclusion in Eq.~(\ref{eq:BC:uR}, \ref{eq:BC:uF}). The red boxes mark the blocks with $i \ne j$ that account for the  interactions between different inclusions. The effect of external stresses is contained in vectors $\mbf{b}_i^\text{disp}$. See  text for  detailed description of elements represented in this system of equations.}
  \label{Fig:LinEqun}
\end{figure}

The boundary conditions in Eq.~(\ref{eq:BC}) have to be satisfied at every point ($\varphi_i$) on the circumference of the $i^\text{th}$ inclusion. Thus the coefficients of the Fourier modes $\{1, \cos(n \varphi_i), \sin (n \varphi_i)\}$ have to match in the expansions of tractions and displacements in Eq.~(\ref{eq:BoundaryTractionsDisplacements}), similar to what was done for the case with the single inclusion in Section~\ref{sec:induction}. This enables us to construct a matrix equation for the set of amplitudes $\{\mbf{a}_{i,\textrm{out}}, \mbf{a}_{i,\textrm{in}}\}$ of induced multipoles  in the form (see also Fig.~\ref{Fig:LinEqun})
 \begin{equation}
     \begin{pmatrix}
     \mbf{M}_{\text{out},ij}^\text{trac}, & \mbf{M}_{\text{in},ij}^\text{trac} \\
    \mbf{M}_{\text{out},ij}^\text{disp}, & \mbf{M}_{\text{in},ij}^\text{disp} \\
     \end{pmatrix}
     \begin{pmatrix}
     \mbf{a}_{j,\textrm{out}} \\
     \mbf{a}_{j,\textrm{in}} \\
     \end{pmatrix}
     =
     \begin{pmatrix}
     \mbf{0} \\
     \mbf{b}_i^\text{disp}\\
     \end{pmatrix},
     \label{eq:matrix}
 \end{equation}
where the summation over inclusions $j$ is implied.
The top and bottom rows of the matrix $\mbf{M}$ in the above equation are obtained from the boundary conditions in Eq.~(\ref{eq:BC}) for tractions (superscript `trac') and displacements (superscript `disp'), respectively. The left and right columns of the matrix $\mbf{M}$ describe the effect of the induced multipoles $\mbf{a}_{i,\textrm{out}}$ and $\mbf{a}_{j,\textrm{in}}$, respectively.  The entries in matrices $\mbf{M}_{\text{out},ii}^\text{trac}$ and $\mbf{M}_{\text{in},ii}^\text{trac}$ for the $i^\text{th}$ inclusion are numbers that depend on the degrees of induced multipoles. 
The entries in matrices $\mbf{M}_{\text{out},ii}^\text{disp}$ and $\mbf{M}_{\text{in},ii}^\text{disp}$ for the $i^\text{th}$ inclusion depend on the degrees of induced multipoles, the radius of inclusion $R_i$ and the material properties of the inclusion~($\mu_i,\kappa_i$) and elastic matrix~($\mu_0,\kappa_0$). Matrices $\mbf{M}_{\text{out},ij}^\text{trac}$ and $\mbf{M}_{\text{out},ij}^\text{disp}$ encode interactions between the inclusions $i$ and $j$. The entries in these matrices depend on the degrees of induced multipoles, the radii $R_i$ and $R_j$ of inclusions, the angle $\theta_{ij}$ and the separation distance $a_{ij}$ between the inclusions (see Fig.~\ref{Fig:ElasMultIllus}). In addition to that, the entries in matrix $\mbf{M}_{\text{out},ij}^\text{disp}$ also depend on the material properties of the elastic matrix ($\mu_0,\kappa_0$). Note that the other matrices are zero, i.e. $\mbf{M}_{\text{in},ij}^\text{trac}=\mbf{M}_{\text{in},ij}^\text{disp}=0$. The entries in vector $\mbf{b}_i^\text{disp}$ depend on the magnitude of external stresses ($\sigma_{xx}^\text{ext}$, $\sigma_{yy}^\text{ext}$, $\sigma_{xy}^\text{ext}$), the degrees of induced multipoles, the radius of inclusion $R_i$, and the material properties of the inclusion ($\mu_i,\kappa_i$) and elastic matrix ($\mu_0,\kappa_0$). Note that in $\mbf{b}_i^\text{disp}$ the only nonzero entries are the ones that correspond to Fourier modes $1$, $\cos(2 \varphi_i)$, and  $\sin(2 \varphi_i)$.

In order to numerically solve the system of equations for induced multipoles in Eq.~(\ref{eq:matrix}) we truncate the multipole expansion at degree $n_\text{max}$. For each inclusion $i$, there are $4 n_\text{max}-1$ unknown amplitudes of multipoles  $\mbf{a}_{i,\text{out}}=\{A_{i,0}, A_{i,1},\dots, A_{i,n_\text{max}}, B_{i,1}, B_{i,2},\dots, B_{i,n_\text{max}}, C_{i,2}, C_{i,3},\dots, C_{i,n_\text{max}}, D_{i,2}, D_{i,3},\dots, D_{i,n_\text{max}} \}$ and  $4 n_\text{max}-1$ unknown amplitudes of multipoles $\mbf{a}_{i,\text{in}}=\{a_{i,2}, a_{i,3},\dots,a_{i,n_\text{max}},b_{i,2}, b_{i,3},\dots,b_{i,n_\text{max}}, c_{i,0}, c_{i,1},\dots,c_{i,n_\text{max}}, d_{i,1}, d_{i,2},\dots, d_{i,n_\text{max}}\}$. Furthermore, we truncate the Taylor expansion for the Airy stress function $\chi_\text{out}\big(r_j(r_i, \varphi_i), \varphi_j(r_i, \varphi_i)|\mbf{a}_{j,\textrm{out}}\big)$ in Eq.~(\ref{eq:AiryOutTaylorExpansion}) at the same order $n_\text{max}$. By matching the coefficients of the Fourier modes $\{1$,  $\cos\varphi_i$, $\sin \varphi_i$, $\ldots$ , $\cos(n_\text{max} \varphi_i)$, $\sin(n_\text{max} \varphi_i)\}$ in the expansions for tractions and displacements in Eq.~(\ref{eq:BoundaryTractionsDisplacements}) around the circumference of the $i^\text{th}$ inclusion, we in principle get $2(2n_\text{max}+1)$ equations from tractions and $2(2n_\text{max}+1)$ equations from displacements. However, the zero Fourier modes for $\sigma_{r\varphi}$ and $u_\varphi$ are equal to zero. Furthermore, the coefficients of Fourier modes $\cos \varphi_i$ and $\sin \varphi_i$ are identical for each of the $\sigma_{rr}$, $\sigma_{r\varphi}$, $u_r$, and $u_\varphi$ in Eq.~(\ref{eq:BoundaryTractionsDisplacements}). By removing the equations that do not provide  new information, the dimensions of matrices $\mathbf{M}_{\text{out},ij}^{\text{trac}}$, $\mathbf{M}_{\text{in},ij}^{\text{trac}}$, $\mathbf{M}_{\text{out},ij}^{\text{disp}}$, and $\mathbf{M}_{\text{in},ij}^{\text{disp}}$, become $(4 n_\text{max}-1)\times(4 n_\text{max}-1)$. Thus Eq.~(\ref{eq:matrix}) describes the system of $N(8 n_\text{max} -2)$ equations for the amplitudes of the induced multipoles $\{\mbf{a}_{1,\text{out}}, \mbf{a}_{1,\text{in}},\ldots,\mbf{a}_{N,\text{out}}, \mbf{a}_{N,\text{in}}\}$. The solution of this system of equations gives amplitudes of induced multipoles, which are linear functions of applied loads $\sigma_{xx}^\text{ext}$, $\sigma_{yy}^\text{ext}$, and $\sigma_{xy}^\text{ext}$.
These amplitudes are then used to obtain the Airy stress functions $\chi^{\text{tot}}_{\text{in}}\big(x,y|\mbf{a}_{i,\text{in}}\big)$ in Eq.~(\ref{eq:AiryIn}) inside inclusions and $\chi^{\text{tot}}_{\text{out}}\big(x,y|\mbf{a}_{\text{out}}\big)$ in Eq.~(\ref{eq:AiryOut}) outside inclusions, which enables us to calculate stresses and displacements everywhere in the structure. The accuracy of the obtained results depends on the number $n_\text{max}$ for the maximum degree of induced multipoles, where larger  $n_\text{max}$ yields more accurate results. In the next two Sections, we compare the results of the elastic multipole method described above with linear finite element simulations and experiments.

\subsection{Comparison with linear finite element simulations}\label{sec:ExampleInfinite}

\begin{figure}[!t]
  \centering
    \includegraphics[scale=1]{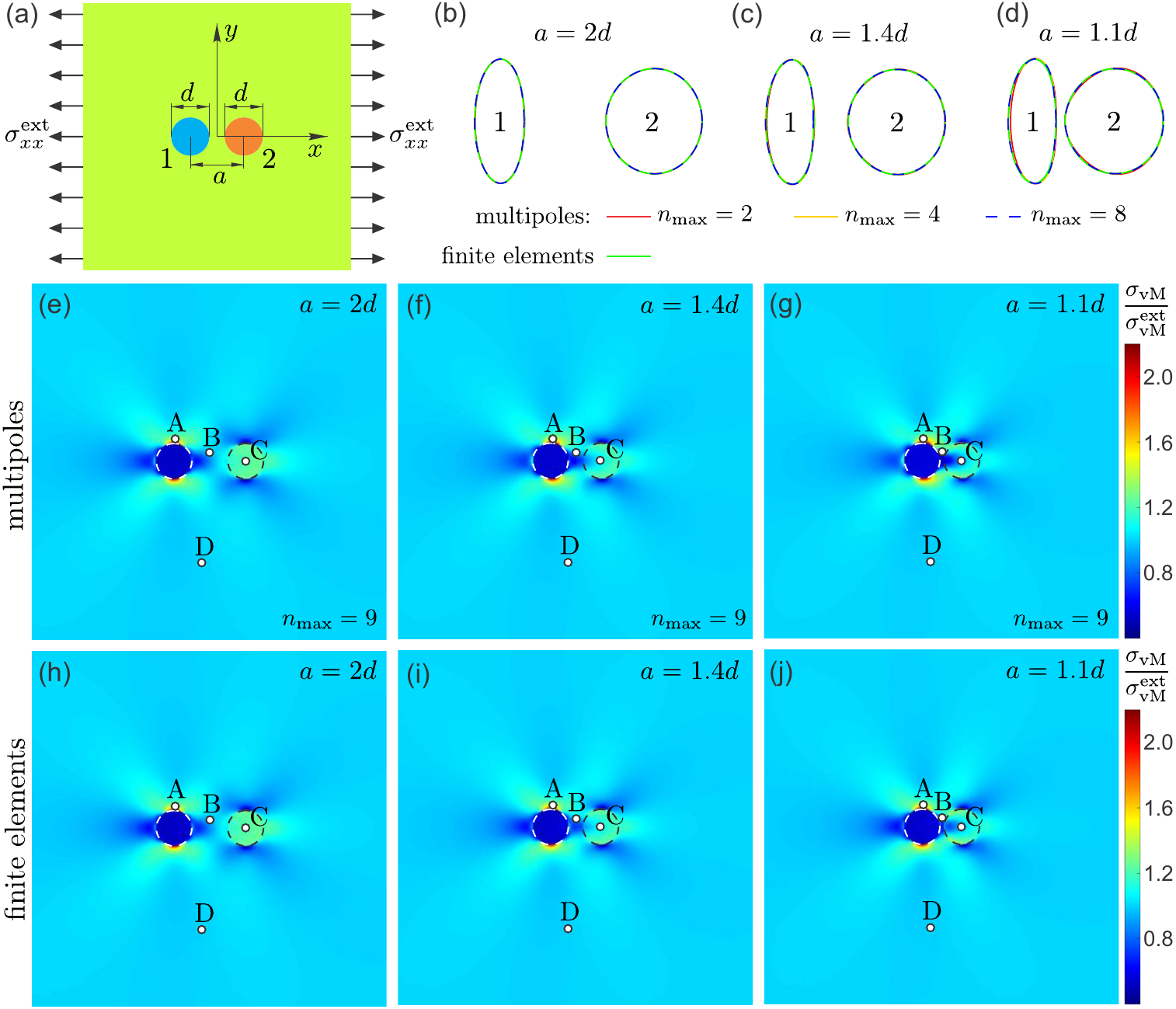}
 \caption{Deformation of an infinite elastic plate with two circular inclusions under uniaxial stress $\sigma_{xx}^\text{ext}$ and plane stress condition. 
(a)~Schematic image describing the initial undeformed shape of the structure and applied load $\sigma_{xx}^\text{ext}= -0.25 E_0$.
The diameters of both inclusions (blue and orange disks) are $d$ and the separation distance between their centers is $a$.  The Young's moduli of the left  and right inclusions  are $E_1/E_0=0.25$ and $E_2/E_0=4$, respectively, where $E_0$ is the Young's modulus of the outer material. Poisson's ratios of the left and right inclusions and the outer material are $\nu_1=0.45$, $\nu_2=0.15$, and $\nu_0=0.3$, respectively. 
(b-d)~Contours of the deformed inclusions for different values of the separation distance $a/d=2$, $1.4$, and $1.1$. The solid red, yellow and dashed blue lines show the contours obtained with elastic multipole method for $n_\text{max}=2$, $4$, and $8$, respectively. Green solid lines show the contours obtained with linear  finite element simulations.
(e-j)~von Mises stress ($\sigma_\text{vM}$) distributions obtained with (e-g)~elastic multipole method ($n_\text{max}=9$) and (h-j)~linear finite element simulations for different separation distances of inclusions $a/d$.  von Mises stress distributions are normalized with the value of von Mises stress $\sigma_\text{vM}^\text{ext}=|\sigma_{xx}^\text{ext}|$ due to the applied load. Four marked points A-D were chosen for the quantitative comparison of stresses and displacements  between elastic multipole method and finite element simulations. See  Table~\ref{tab:TwoIncUniaxial} for details. 
}
\label{Fig:TwoIncUniaxial}
\end{figure}

\begin{figure}[!t]
  \centering
  \includegraphics[scale=1]{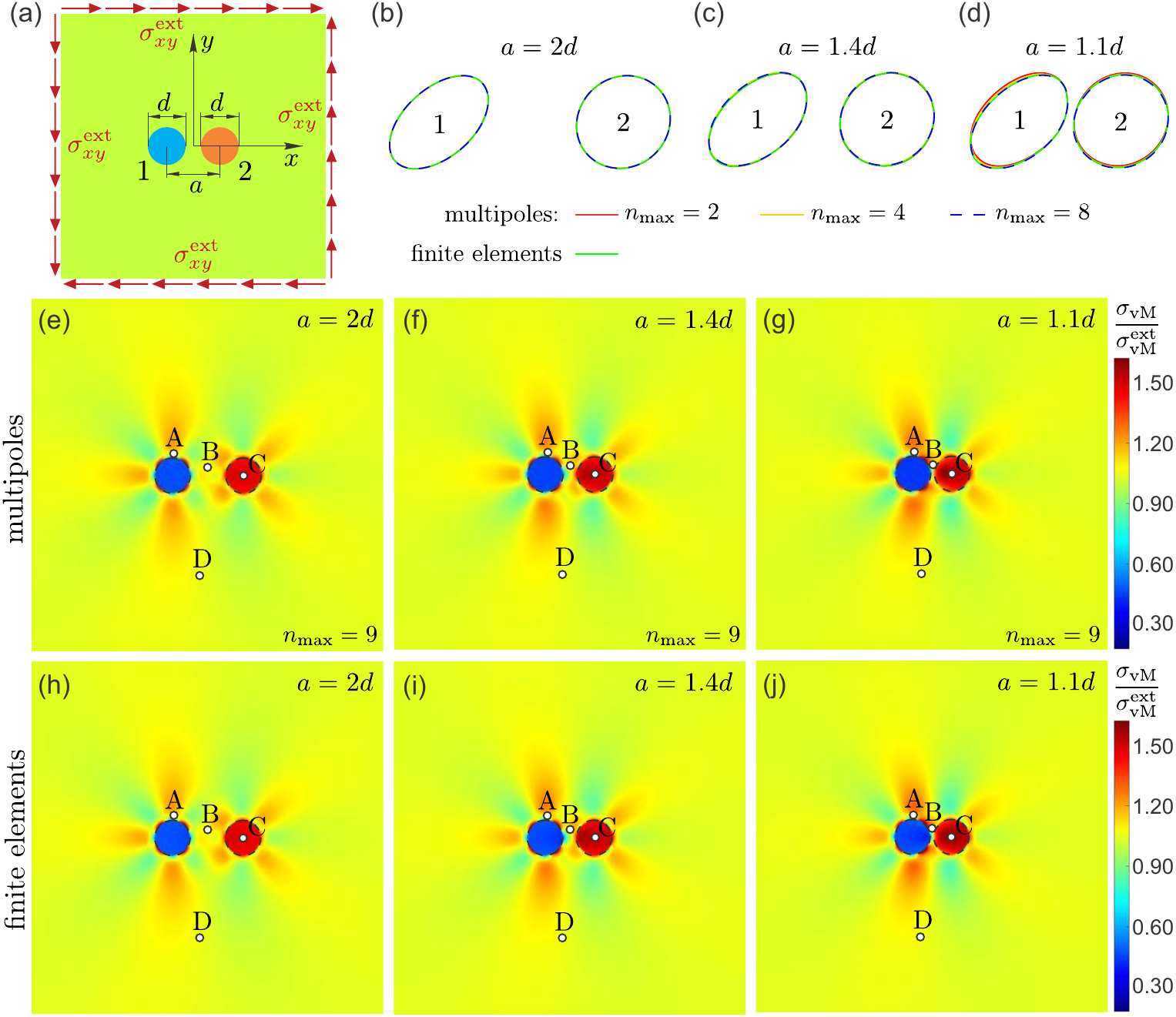}
  \caption{Deformation of an infinite elastic plate with two circular inclusions under shear stress $\sigma_{xy}^\text{ext}$ and plane stress condition.
(a)~Schematic image describing the initial undeformed shape of the structure and  applied load $\sigma_{xy}^\text{ext}=0.1 E_0$, where $E_0$ is the Young's modulus of the outer material. 
The diameter of both inclusions (blue and orange disks) is $d$ and the separation distance between their centers is $a$. Material properties are the same as in Fig.~\ref{Fig:TwoIncUniaxial}. (b-d)~Contours of the deformed inclusions for different values of the separation distance $a/d=2$, $1.4$, and $1.1$. The solid red, yellow and dashed blue lines show the contours obtained with elastic multipole method for $n_\text{max}=2$, $4$, and $8$, respectively. Green solid lines show the contours obtained with finite element simulations.
(e-j)~von Mises stress ($\sigma_\text{vM}$) distributions obtained with (e-g)~elastic multipole method ($n_\text{max}=9$) and (h-j)~linear finite element simulations for different separation distances of inclusions $a/d$.  von Mises stress distributions are normalized with the value of von Mises stress $\sigma_\text{vM}^\text{ext}=\sqrt{3}|\sigma_{xy}^\text{ext}|$ due to the applied load. Four marked points A-D were chosen for the quantitative comparison of stresses and displacements  between elastic multipole method and  finite element simulation. See Table~\ref{tab:TwoIncShear} for details.
}
\label{Fig:TwoIncShear}
\end{figure}

First, we tested the elastic multipole method for two circular inclusions embedded in an infinite plate subjected to uniaxial stress (Fig.~\ref{Fig:TwoIncUniaxial}) and shear stress (Fig.~\ref{Fig:TwoIncShear}). The two inclusions had identical diameters $d$ and they were centered at $(\pm a/2,0)$.  Three different values of the separation distance $a$ between the inclusions were considered: $a=2d$, $a=1.4d$, and $a=1.1d$.
The left and right inclusions were chosen to be more flexible ($E_1/E_0=0.25$) and stiffer ($E_2/E_0=4$) than the outer matrix with the Young's modulus $E_0$, respectively. We used plane stress condition with Kolosov's constants $\kappa_i=(3-\nu_i)/(1+\nu_i)$, where Poisson's ratios of the left and right inclusions, and the outer material were $\nu_1=0.45$, $\nu_2=0.15$, and $\nu_0=0.3$, respectively.
The values of the applied uniaxial stress and shear stress were
$\sigma_{xx}^{\text{ext}}/E_0 = -0.25$ (Fig.~\ref{Fig:TwoIncUniaxial}) and $\sigma_{xy}^{\text{ext}}/E_0 = 0.1$ (Fig.~\ref{Fig:TwoIncShear}), respectively. Such large values of external loads were used only to exaggerate deformations. Note that in practical experiments these loads would cause nonlinear deformation. 

In Figs.~\ref{Fig:TwoIncUniaxial} and~\ref{Fig:TwoIncShear} we show contours of deformed inclusions and spatial distributions of stresses for different values of the separation distance $a$ between the inclusions, where the results from elastic multipole method were compared with linear finite element simulations on a square domain of size $400d \times 400d$ (see Appendix~\ref{app:FEM} for details). When the inclusions are far apart, they interact weakly, as can be seen from the expansion of stresses and displacements in Eq.~(\ref{eq:BoundaryTractionsDisplacements}), where the terms describing interactions between the inclusions $i$ and $j$  contain powers of $R_i/a_{ij}\ll 1$ and $R_j/a_{ij} \ll 1$. This is the case for the separation distance $a=2d$, where we find that the contours of deformed inclusions have elliptical shapes (see Figs.~\ref{Fig:TwoIncUniaxial}b and~\ref{Fig:TwoIncShear}b) and stresses inside the inclusions are uniform (see Figs.~\ref{Fig:TwoIncUniaxial}e,h and~\ref{Fig:TwoIncShear}e,h), which is characteristic for isolated inclusions (see Eq.~(\ref{eq:SingleInclusionStress}) and \cite{Eshelby}). Furthermore, the von Mises stress distribution ($\sigma_\text{vM}=\sqrt{\sigma_{xx}^2-\sigma_{xx}\sigma_{yy}+\sigma_{yy}^2+3\sigma_{xy}^2}$) around the more flexible left inclusion (see Fig.~\ref{Fig:TwoIncUniaxial}e,h) is similar to that of an isolated hole under uniaxial stress (see Fig.~\ref{fig:induction}c). For the stiffer right inclusion, the locations of the maxima and minima in the von Mises stress distribution are reversed (see Fig.~\ref{Fig:TwoIncUniaxial}e,h) because the amplitudes of  induced multipoles have the opposite sign (see Eq.~(\ref{eq:ElasInductioninclusion})).

Similar patterns in the von Mises stress distribution are observed when the structure is under external shear, but they are rotated by 45$^\circ$ (see Fig.~\ref{Fig:TwoIncShear}e,h).
When inclusions are far apart ($a=2d$), the contours of deformed inclusions can  be accurately described already with multipoles up to degree $n_\text{max}=2$ (see Figs.~\ref{Fig:TwoIncUniaxial}b and~\ref{Fig:TwoIncShear}b). This degree of multipoles is sufficient because external stresses $\sigma_{xx}^{\text{ext}}$ and $\sigma_{xy}^{\text{ext}}$ couple only to the Fourier modes $1$, $\cos 2 \varphi_i$, and $\sin 2 \varphi_i$ in the expansion for stresses and displacements in Eq.~(\ref{eq:BoundaryTractionsDisplacements}). 
As the inclusions are moved closer together ($a=1.4d$ and $a=1.1d$), they  interact more strongly. As a consequence, the contours of deformed inclusions become  progressively more non-elliptical and higher order of multipoles are needed to accurately describe their shapes (see Figs.~\ref{Fig:TwoIncUniaxial}c,d and~\ref{Fig:TwoIncShear}c,d). Furthermore, the stress distribution inside the right inclusion becomes nonuniform (see Figs.~\ref{Fig:TwoIncUniaxial}f,g,i,j and~\ref{Fig:TwoIncShear}f,g,i,j). Note that von Mises stress distributions  look similar far from inclusions regardless of the separation distance $a$ (see Figs.~\ref{Fig:TwoIncUniaxial} and~\ref{Fig:TwoIncShear}), because they are dictated by the lowest order induced multipoles, i.e. by non-topological monopoles ($p$), non-topological dipoles ($\mbf{d}^p$) and quadrupoles ($\mbf{Q}^s$, $\mbf{Q}^p$).

\begin{table}[!t]
\centering
\caption{Quantitative comparison for the values of von Mises stresses $\sigma_\text{vM}$ and displacements $\mbf{u}$ at  points A-D (defined in Fig.~\ref{Fig:TwoIncUniaxial}) obtained with the elastic multipole method (EMP) and finite element simulations (FEM) for uniaxially compressed samples with two inclusions of diameter $d$ for different values of their separation distance $a$. von Mises stresses $\sigma_\text{vM}$ are normalized with the value of von Mises stress $\sigma_\text{vM}^\text{ext}=|\sigma_{xx}^\text{ext}|$ due to the applied uniaxial compression. Displacements $\mbf{u}$ are normalized with the characteristic scale of deformation $d\sigma_\text{vM}^\text{ext}/E_0$, where $E_0$ is the Young's modulus of the elastic matrix.  The relative percent errors $\epsilon$ between the two methods are calculated as $100\times(\sigma_\text{vM}^\text{EMP}-\sigma_\text{vM}^\text{FEM})/\sigma_\text{vM}^\text{FEM}$ and $100\times(|\mbf{u}|^\text{EMP}-|\mbf{u}|^\text{FEM})/|\mbf{u}|^\text{FEM}$.}
\label{tab:TwoIncUniaxial}
\def\arraystretch{1.15}
\begin{tabular}{|c|ccc|>{\centering}p{0.0475\textwidth}>{\centering}p{0.0475\textwidth}>{\centering}p{0.045\textwidth}|ccc|>{\centering}p{0.0475\textwidth}>{\centering}p{0.0475\textwidth}>{\centering}p{0.045\textwidth}|ccc|>{\centering}p{0.0475\textwidth}>{\centering}p{0.0475\textwidth}>{\centering\arraybackslash}p{0.045\textwidth}|}
\hline
& \multicolumn{6}{c|}{separation $a=2d$}& \multicolumn{6}{c|}{separation $a=1.4d$}& \multicolumn{6}{c|}{separation $a=1.1d$}\\
\cline{2-19}
& \multicolumn{3}{c|}{stress $\sigma_\text{vM}/\sigma_\text{vM}^\text{ext}$} & \multicolumn{3}{c|}{disp.  $|\mathbf{u}|/[d\sigma_\text{vM}^\text{ext}/E_0]$} & \multicolumn{3}{c|}{stress $\sigma_\text{vM}/\sigma_\text{vM}^\text{ext}$} & \multicolumn{3}{c|}{disp.  $|\mathbf{u}|/[d\sigma_\text{vM}^\text{ext}/E_0]$} & \multicolumn{3}{c|}{stress $\sigma_\text{vM}/\sigma_\text{vM}^\text{ext}$} & \multicolumn{3}{c|}{disp.  $|\mathbf{u}|/[d\sigma_\text{vM}^\text{ext}/E_0]$} \\
\cline{2-19}
  &  EMP   & FEM   & $\epsilon$ $(\%)$  &  EMP   & FEM   & $\epsilon(\%)$ &  EMP   & FEM   & $\epsilon$ $(\%)$ &  EMP   & FEM   & $\epsilon(\%)$ &  EMP   & FEM   & $\epsilon(\%)$ &  EMP   & FEM   & $\epsilon(\%)$\\
\hline
A & 1.419 & 1.416 & 0.2 & 1.442 & 1.442  & 0.0 & 1.424 & 1.419 & 0.5 & 1.265 & 1.264 & 0.1 & 1.439 & 1.426 & 0.9 & 1.223 & 1.220   & 0.3 \\
B & 0.940   & 0.947 & 0.7 & 0.082 & 0.081 & 0.5 & 0.940 & 0.959 & 1.9 & 0.116 & 0.116 & 0.7 & 0.887 & 0.91  & 2.7 & 0.127 & 0.126 & 1.0 \\
C & 1.213 & 1.216 & 0.3 & 0.654 & 0.653 & 0.1 & 1.083 & 1.092 & 0.8 & 0.275 & 0.274 & 0.3 & 0.933 & 0.948 & 1.5 & 0.122  & 0.122 & 0.3 \\
D & 0.997 & 0.997 & 0.0 & 1.144  & 1.144  & 0.0 & 0.994 & 0.994 & 0.0 & 1.250   & 1.250   & 0.0 & 0.992  & 0.992  & 0.0 & 1.363 & 1.363 & 0.0\\
\hline            
\end{tabular}
\end{table}

\begin{table}[!t]
\centering
\caption{Quantitative comparison for the values of von Mises stresses $\sigma_\text{vM}$ and displacements $\mbf{u}$ at  points A-D (defined in Fig.~\ref{Fig:TwoIncShear}) obtained with the elastic multipole method (EMP) and finite element simulations (FEM) for sheared samples with two inclusions of diameter $d$ for different values of their separation distance $a$. von Mises stresses $\sigma_\text{vM}$ are normalized with the value of von Mises stress $\sigma_\text{vM}^\text{ext}=\sqrt{3}|\sigma_{xy}^\text{ext}|$ due to the applied shear. Displacements $\mbf{u}$ are normalized with the characteristic scale of deformation $d\sigma_\text{vM}^\text{ext}/E_0$, where $E_0$ is the Young's modulus of the elastic matrix.  The relative percent errors $\epsilon$ between the two methods are calculated as $100\times(\sigma_\text{vM}^\text{EMP}-\sigma_\text{vM}^\text{FEM})/\sigma_\text{vM}^\text{FEM}$ and $100\times(|\mbf{u}|^\text{EMP}-|\mbf{u}|^\text{FEM})/|\mbf{u}|^\text{FEM}$.}
\label{tab:TwoIncShear}
\def\arraystretch{1.15}
\begin{tabular}{|c|ccc|>{\centering}p{0.0475\textwidth}>{\centering}p{0.0475\textwidth}>{\centering}p{0.045\textwidth}|ccc|>{\centering}p{0.0475\textwidth}>{\centering}p{0.0475\textwidth}>{\centering}p{0.045\textwidth}|ccc|>{\centering}p{0.0475\textwidth}>{\centering}p{0.0475\textwidth}>{\centering\arraybackslash}p{0.045\textwidth}|}
\hline
& \multicolumn{6}{c|}{separation $a=2d$}& \multicolumn{6}{c|}{separation $a=1.4d$}& \multicolumn{6}{c|}{separation $a=1.1d$}\\
\cline{2-19}
& \multicolumn{3}{c|}{stress $\sigma_\text{vM}/\sigma_\text{vM}^\text{ext}$} & \multicolumn{3}{c|}{disp.  $|\mathbf{u}|/[d\sigma_\text{vM}^\text{ext}/E_0]$} & \multicolumn{3}{c|}{stress $\sigma_\text{vM}/\sigma_\text{vM}^\text{ext}$} & \multicolumn{3}{c|}{disp.  $|\mathbf{u}|/[d\sigma_\text{vM}^\text{ext}/E_0]$} & \multicolumn{3}{c|}{stress $\sigma_\text{vM}/\sigma_\text{vM}^\text{ext}$} & \multicolumn{3}{c|}{disp.  $|\mathbf{u}|/[d\sigma_\text{vM}^\text{ext}/E_0]$} \\
\cline{2-19}
  &  EMP   & FEM   & $\epsilon$ $(\%)$  &  EMP   & FEM   & $\epsilon$ $(\%)$ &  EMP   & FEM   & $\epsilon$ $(\%)$ &  EMP   & FEM   & $\epsilon$ $(\%)$ &  EMP   & FEM   & $\epsilon$ $(\%)$ &  EMP   & FEM   & $\epsilon$ $(\%)$\\
\hline
A & 1.027 & 1.026 & 0.1 & 1.167 & 1.166 & 0.1 & 1.051  & 1.047 & 0.4 & 1.132 & 1.131 & 0.1 & 1.085 & 1.076 & 0.9 & 1.159  & 1.156 & 0.2 \\
B & 1.045 & 1.050  & 0.5 & 0.387 & 0.387  & 0.2 & 1.065 & 1.081 & 1.5 & 0.379 & 0.377 & 0.5 & 1.106 & 1.134 & 2.5 & 0.371 & 0.368 & 0.8 \\
C & 1.415  & 1.419 & 0.2 & 0.159 & 0.159 & 0.0 & 1.469 & 1.481 & 0.8 & 0.280 & 0.279  & 0.2 & 1.512 & 1.533 & 1.4 & 0.407 & 0.405 & 0.4 \\
D & 1.016 & 1.016 & 0.0 & 4.235 & 4.235 & 0.0 & 1.016 & 1.016 & 0.0 & 4.265 & 4.265 & 0.0 & 1.020 & 1.020 & 0.0 & 4.275 & 4.275 & 0.0\\
\hline            
\end{tabular}
\end{table}

\begin{figure}[!t]
\centering
    \includegraphics[scale=1]{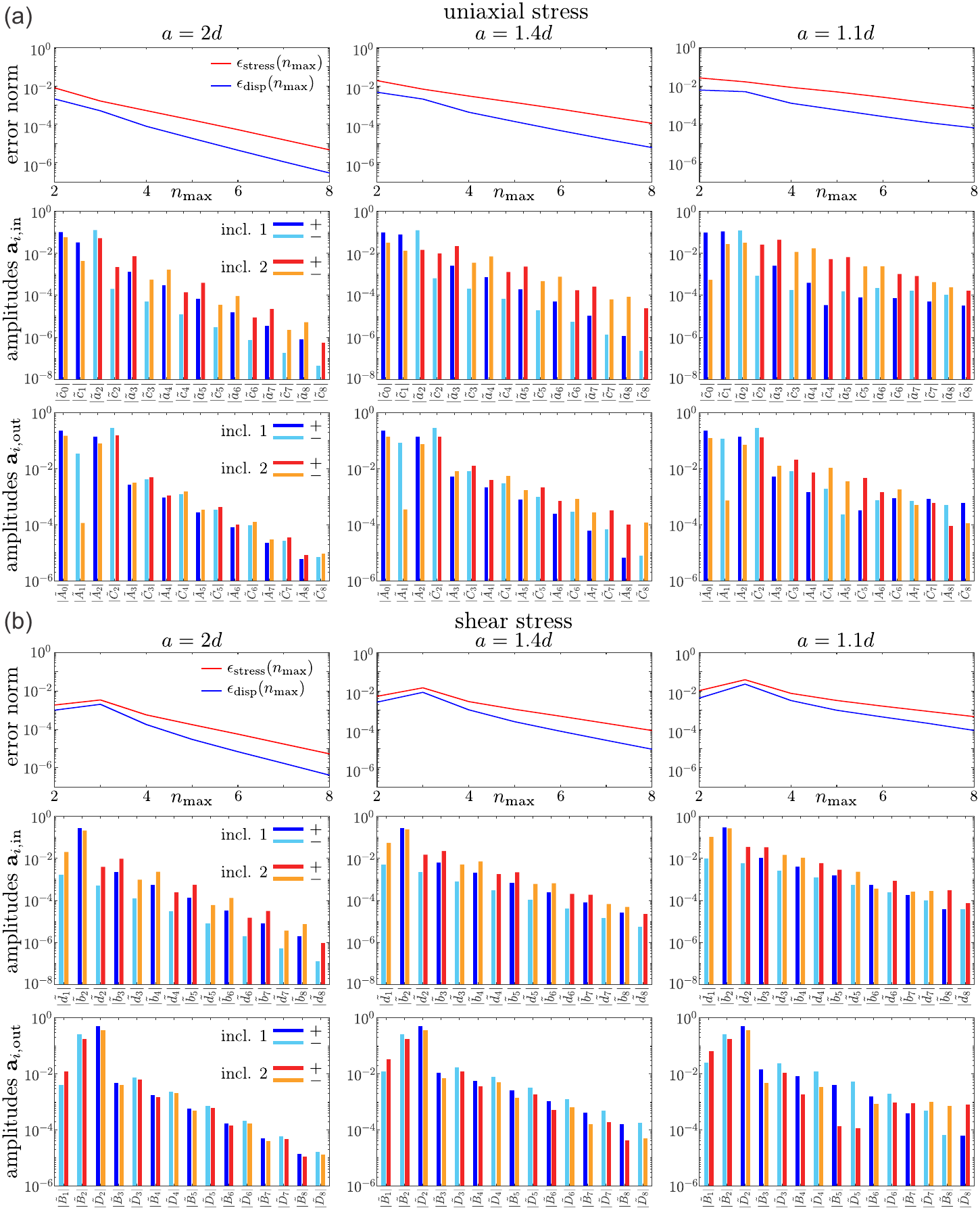}
\caption{Normalized errors and amplitudes of induced multipoles for the structures with two inclusions with diameters $d$ and the separation distance $a$ under (a)~uniaxial stress $\sigma_{xx}^\text{ext}$ (see Fig.~\ref{Fig:TwoIncUniaxial}) and (b) shear stress $\sigma_{xy}^\text{ext}$ (see Fig.~\ref{Fig:TwoIncShear}). The normalized errors for displacements $\epsilon_\text{disp}(n_\text{max})$ (blue lines) and stresses $\epsilon_\text{stress}(n_\text{max})$ (red lines) are defined in Eq.~(\ref{eq:error}). Absolute values of the amplitudes of induced multipoles $\{\mbf{a}_{1,\text{out}},\mbf{a}_{2,\text{out}}\}$ for $n_\text{max}=9$. In (a) the amplitudes are normalized as $\tilde{a}_n=a_n/\sigma_{xx}^\text{ext}$, $\tilde{c}_n=c_n/\sigma_{xx}^\text{ext}$, $\tilde{A}_n=A_n/\sigma_{xx}^\text{ext}$, and $\tilde{C}_n=C_n/\sigma_{xx}^\text{ext}$. The dark and light blue colored bars correspond to the positive ($a_n, c_n, A_n, C_n>0$) and negative ($a_n, c_n,A_n,C_n<0$) amplitudes for inclusion 1, respectively. Similarly, the red and orange colored bars correspond to the positive and negative amplitudes for inclusion 2, respectively. Note that the amplitudes of multipoles $b_i$, $d_i$, $B_i$, and $D_i$ are zero due to the symmetry of the problem. In (b)~the amplitudes are normalized as $\tilde{b}_n=b_n/\sigma_{xy}^\text{ext}$, $\tilde{d}_n=d_n/\sigma_{xy}^\text{ext}$, $\tilde{B}_n=B_n/\sigma_{xy}^\text{ext}$, and $\tilde{D}_n=D_n/\sigma_{xy}^\text{ext}$. The dark and light blue colored bars correspond to the positive ($b_n, d_n, B_n, D_n>0$) and negative ($b_n, d_n,B_n,D_n<0$) amplitudes for inclusion 1, respectively. Similarly, the red and orange colored bars correspond to the positive and negative amplitudes for inclusion 2, respectively. Note that the amplitudes of multipoles $a_i$, $c_i$, $A_i$, and $C_i$ are zero due to the symmetry of the problem. }
\label{Fig:convAmp}
\end{figure}

To determine the proper number for the maximum degree $n_\text{max}$ of induced multipoles we performed a convergence analysis for the spatial distributions of displacements $\mbf{u}^{(n_\text{max})}(x,y)$ and von Mises stresses $\sigma^{(n_\text{max})}_\text{vM}(x,y)$. Displacements and von Mises stresses were evaluated at $N_p=1001 \times 1001$ points on a square grid of size $10d \times 10d$ surrounding the inclusions, i.e. at the points $(x_i,y_j)=\left(id/100,jd/100\right)$, where $i,j\in\{-500,-499,\ldots,500\}$. The normalized errors for displacements $\epsilon_\text{disp}(n_\text{max})$ and stresses $\epsilon_\text{stress}(n_\text{max})$ were obtained by calculating the relative changes of the spatial distributions of displacements and von Mises stresses when the maximum degree $n_\text{max}$ of induced multipoles is increased by one. The normalized errors are given by~\cite{SpecMethod}
\begin{subequations}
\begin{align}
\label{Eqn:errordisp}
    \epsilon_\text{disp}(n_\text{max}) &= \frac{1}{\sqrt{N_p}}\left[\sum_{i,j}\left|\left(\frac{\mbf{u}^{(n_\text{max}+1)}(x_i,y_j)-\mbf{u}^{(n_\text{max})}(x_i,y_j)}{d\,\sigma_\text{vM}^\text{ext}/E_0}\right)^2 \right|\right]^{1/2},\\
\label{Eqn:errorstress}
    \epsilon_\text{stress}(n_\text{max}) &= \frac{1}{\sqrt{N_p}}\left[\sum_{i,j}\left(\frac{\sigma^{(n_\text{max}+1)}_\text{vM}(x_i,y_j)-\sigma^{(n_\text{max})}_\text{vM}(x_i,y_j)}{\sigma_\text{vM}^\text{ext}}\right)^2 \right]^{1/2}.
\end{align}
\label{eq:error}%
\end{subequations}
Here, displacements and von Mises stresses are normalized by the characteristic scales $d \sigma_\text{vM}^\text{ext}/E_0$ and $\sigma_\text{vM}^\text{ext}$, respectively, where $d$ is the diameter of inclusions, $\sigma_\text{vM}^\text{ext}$ is the value of the von Mises stress due to external load, and $E_0$ is the Young's modulus of the surrounding matrix. The normalized errors are plotted in Fig.~\ref{Fig:convAmp}. As the maximum degree $n_\text{max}$ of induced multipoles is increased, the normalized errors for displacements $\epsilon_\text{disp}(n_\text{max})$ and stresses $\epsilon_\text{stress}(n_\text{max})$ decrease exponentially. Since the
induced elastic multipoles form the basis for the biharmonic equation, this is akin to the spectral method, which is exponentially convergent when the functions and the boundaries are smooth~\cite{SpecMethod}. The normalized errors for displacements are lower than the errors for stresses (see Fig.~\ref{Fig:convAmp}) because stresses are related to spatial derivatives of displacements. Note that the normalized errors decrease more slowly when inclusions are brought close together and their interactions become important (see Fig.~\ref{Fig:convAmp}). This is also reflected in the amplitudes $\mbf{a}_{1,\text{in}}$, $\mbf{a}_{2,\text{in}}$, $\mbf{a}_{1,\text{out}}$, and $\mbf{a}_{2,\text{out}}$ of the induced multipoles, which decrease exponentially with the degree of  multipoles and they decrease more slowly when inclusions are closer  (see Fig.~\ref{Fig:convAmp}).

Results from the elastic multipole method were compared with linear finite element simulations, and very good agreement is achieved already for $n_\text{max}=9$ even when inclusions are very close together ($a=1.1d$, see Figs.~\ref{Fig:TwoIncUniaxial} and~\ref{Fig:TwoIncShear}). To make the comparison with finite elements more quantitative, we compared the values of displacements and stresses at 4 different points: A~--~at the edge of the left inclusion, B~--~in between the inclusions, C~--~at the center of the right inclusion, and  D~--~far away from both inclusions (see Figs.~\ref{Fig:TwoIncUniaxial} and~\ref{Fig:TwoIncShear}). For all 4 points, the error increases when inclusions are brought closer together (see Tables~\ref{tab:TwoIncUniaxial} and \ref{tab:TwoIncShear}). Of the 4 different points, we find that the errors are the largest at point B, which is strongly influenced by induced multipoles from both inclusions. For the smallest separation distance $a=1.1d$ between the inclusions, the errors for the von Mises stress at point B are $2.7\%$ and $2.5\%$ for the uniaxial and shear loads, respectively. These errors can be further reduced by increasing the number $n_\text{max}$ for the maximum degree of multipoles, e.g. for $n_\text{max}=14$ the errors for von Mises stress at point B are reduced to $1.2\%$ and $1.1\%$ for the uniaxial and shear loads, respectively.

\begin{figure}[!t]
  \centering
  \includegraphics[scale=1.]{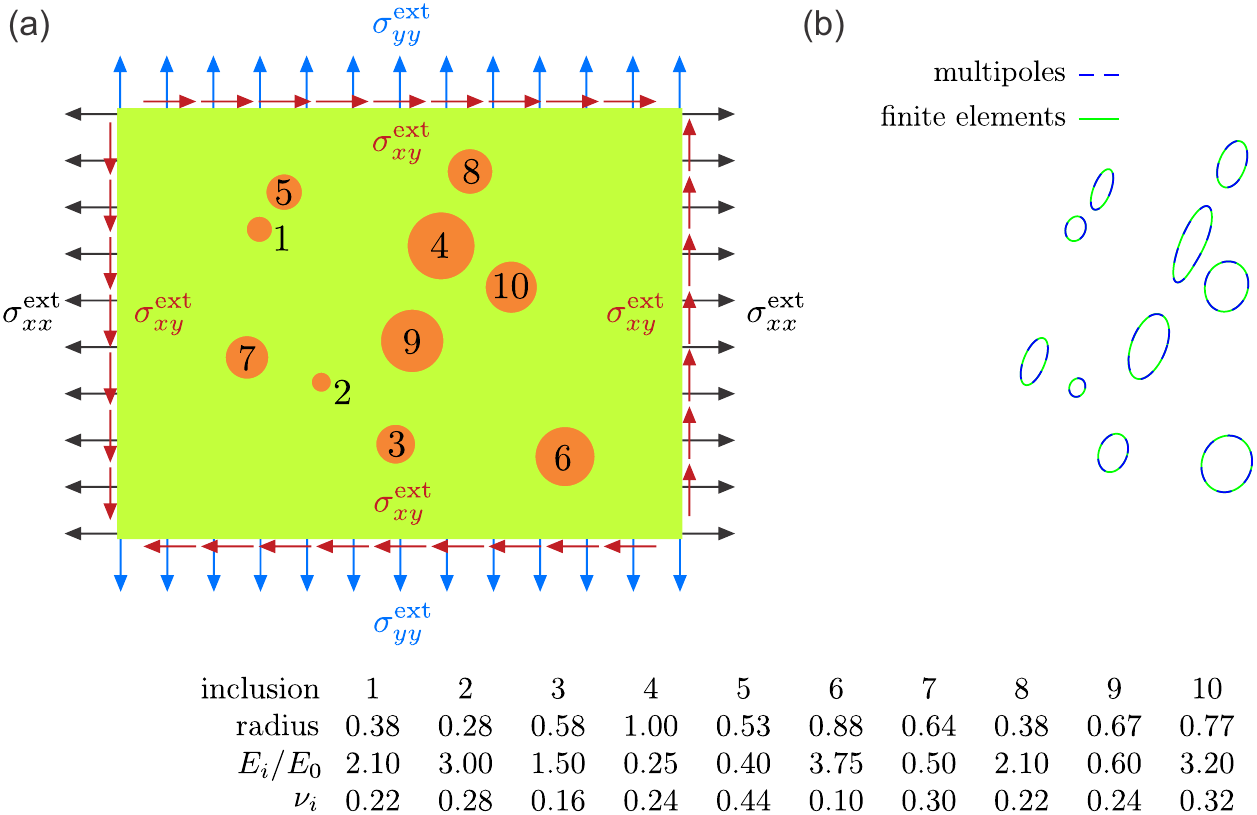}
  \caption{Deformation of an infinite elastic plate with ten circular inclusions (orange disks) under general  external stress. 
    (a)~Schematic image describing the initial undeformed shape of the structure and  applied external loads: $\sigma_{xx}^{\text{ext}}/E_0 = -0.25$, $\sigma_{yy}^{\text{ext}}/E_0 = 0.05$, and $\sigma_{xy}^{\text{ext}}/E_0 = 0.10$, where $E_0$ is the Young's modulus of the outer material. The plane stress condition was used. The radii and material properties (Young's moduli $E_i$ and Poisson's ratios $\nu_i$) of inclusions are provided in the table  below the schematic image. The radii of inclusions are normalized by the radius of the largest inclusion. The Young's moduli are normalized by the Young's modulus of the outer material $E_0$. The value of Poisson's ratio for the outer material was $\nu_0=0.3$.
    (b)~Contours of deformed inclusions. The blue dashed lines show the results obtained with the elastic multipole method ($n_\text{max}=6$). The green solid lines correspond to the deformed contours obtained with linear finite element simulations. 
    }
  \label{Fig:TenIncGen}
\end{figure}
To demonstrate the full potential of the elastic multipole method, we also considered the deformation of an infinite plate   with $N=10$ inclusions of different sizes and material properties subjected to general external stress load under plane stress condition (see 
Fig.~\ref{Fig:TenIncGen}).
The contours of deformed inclusions obtained with finite element simulations (green solid lines) and elastic multipole method with $n_\text{max}=6$ (blue dashed lines) are in very good agreement. Note that the results for the elastic multipole method were obtained by solving the linear system of only $N(8 n_\text{max}-2)=460$ equations for the amplitudes of the induced multipoles described in  Eq.~(\ref{eq:matrix}), which is significantly smaller than the number of degrees of freedom required for finite element simulations. 

\subsection{Comparison with experiments}
\label{sec:Experiments}

Finally, we also tested the elastic multipole method against experiments. Experimental samples were prepared by casting Elite Double 32 (Zhermack) elastomers with the measured Young's modulus $E_0=0.97$~MPa and assumed Poisson's ratio $\nu=0.49$~\cite{babaee20133d}.
Molds were fabricated from 5 mm thick acrylic plates with laser-cut circular holes, which were then filled with acrylic cylinders in the assembled molds to create cylindrical holes in the elastomer samples. Approximately 30~min after casting, the molds were disassembled and the solid samples were placed in a convection oven at 40\;$^\circ$C for 12~hours for further curing. The cylindrical inclusions made from acrylic (Young's modulus $E=2.9$~GPa, Poisson's ratio $\nu=0.37$~\cite{acrylic})
were inserted into the holes of the elastomer samples and they were glued by a cyanoacrylate adhesive. 

\begin{figure}[!t]
\centering
  \includegraphics[width=\textwidth]{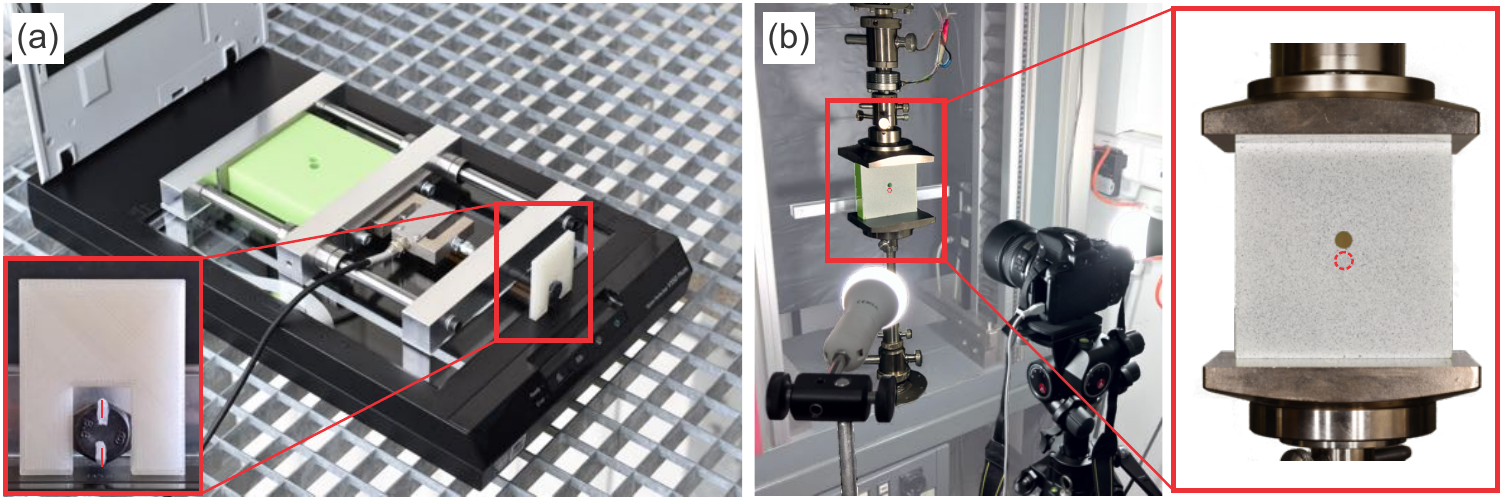}
  \caption{Experimental systems for displacement controlled compressive tests. (a)~A mechanism for compression of rubber samples (green slab) sits on top of a flatbed photo scanner, which is used to extract the contours of deformed holes/inclusions. The zoomed-in photo on the left shows a 3D-printed plastic wrench that was used for the precise control of screw turns. (b)~Setup for extracting strain fields via digital image correlation (DIC). The surface of the sample was painted with speckle patterns. The sample was then compressed with steel plates of the testing machine and photos of speckle patterns were used to extract the displacement field on the front surface of the slab. The zoomed-in photo on the right shows the rubber sample  with one hole and one inclusion (indicated with a red dashed circle) mounted between two parallel plates of the testing machine.}
  \label{Fig:ExperimentalSetup}
\end{figure}
We designed two compressive testing systems (see Fig.~\ref{Fig:ExperimentalSetup}) to compare the contours of deformed holes and inclusions and strain fields with predictions made by the elastic multipole method. In Fig.~\ref{Fig:ExperimentalSetup}a, we first present an experimental system for extracting the contours of deformed holes and inclusions in compressed experimental samples. The system comprises a custom-made loading mechanism and a flatbed photo scanner. Displacement loading is applied in 0.5~mm increments via $180^{\circ}$ turns of the M10x1 screw (metric thread with 10~mm  diameter and 1~mm pitch) in the mechanism, which   is controlled by a 3D-printed plastic wrench (see the inset of Fig.~\ref{Fig:ExperimentalSetup}a).
The loading mechanism was placed on an Epson V550 photo scanner to scan the surface of deformed samples and silicone oil was applied between the sample and the glass surface of the scanner to reduce  friction between them. Scanned images were post-processed with Corel PHOTO-PAINT X8 and the Image Processing Toolbox in MATLAB 2018b. First,  the dust particles and air bubbles trapped in a thin film of silicone oil  were digitally removed from the scanned images. Scanned grayscale images were then converted to black and white binary images from which the contours were obtained with MATLAB. 

Second, we present a system for capturing the displacement and strain fields in compressed samples via  digital image correlation (DIC) technique (see Fig.~\ref{Fig:ExperimentalSetup}b). Black and white speckle patterns were spray-painted onto the surface of  samples  with  slow-drying acrylic paint to prevent the speckle pattern  from hardening too  quickly, which could lead to delamination under applied compressive loads. Using a Zwick Z050 universal material testing machine, we applied  a  compressive displacement in 0.2~mm increments, where  again a silicone oil was applied between the steel plates and the elastomer samples to prevent sticking and to reduce friction. A Nikon D5600 photo camera was used at each step to take a snapshot of the compressed sample (see Fig.~\ref{Fig:ExperimentalSetup}b). These photos were then used to calculate the displacements and strain fields with  Ncorr, an open-source 2D DIC MATLAB based software.~\cite{DIC}

\begin{figure}[!t]
\centering
  \includegraphics[width=\textwidth]{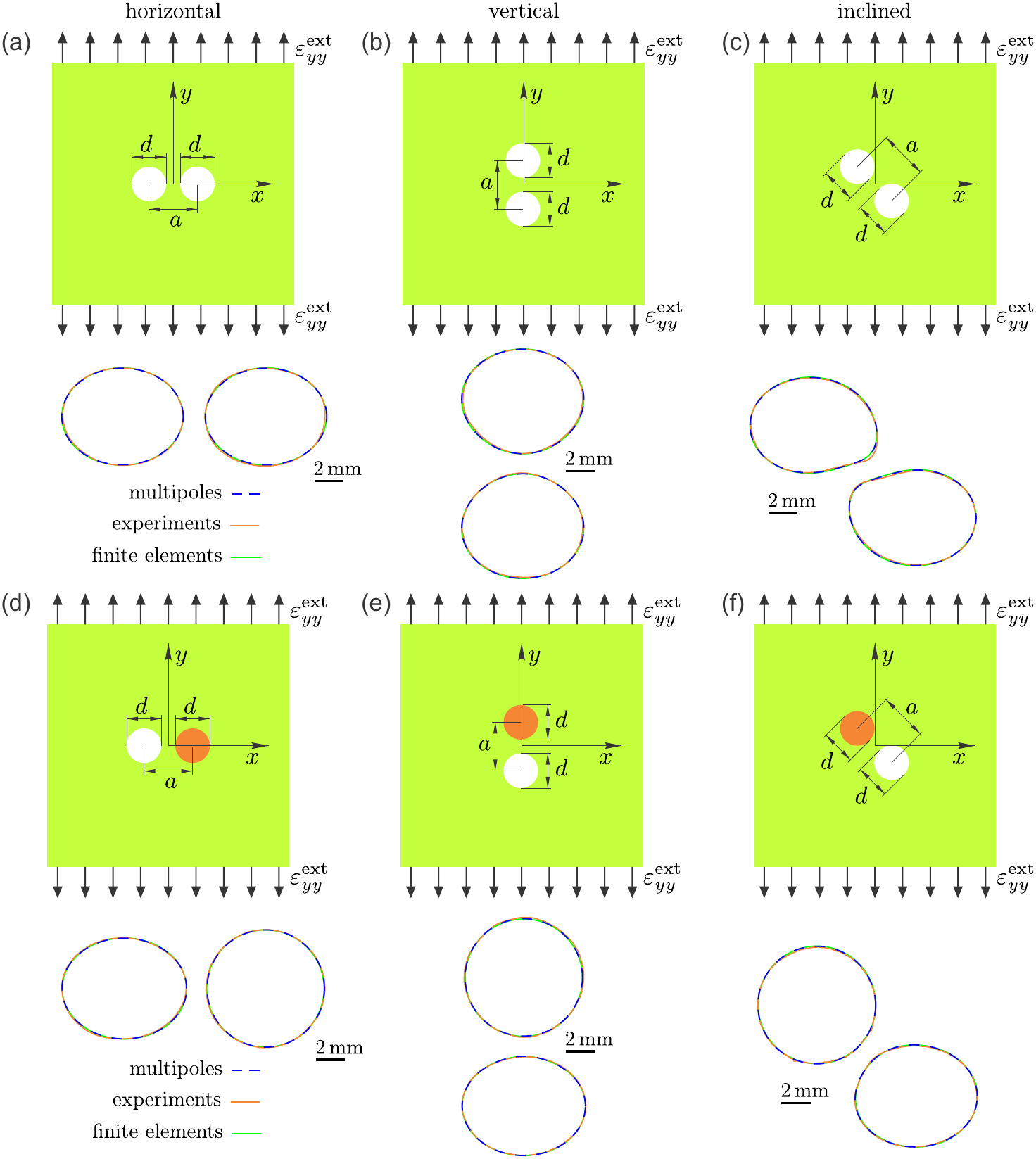}
  \caption{Uniaxial vertical compression of elastic structures with holes and inclusions. (a-c)~Schematic images describing the initial undeformed shapes of structures with two holes (white disks) in three different configurations (horizontal, vertical and inclined at $45^\circ$) and  applied external strain $\varepsilon_{yy}^{\text{ext}}=-0.05$. Deformed contours of holes obtained with elastic multipole method ($n_\text{max}=10$, blue dashed lines), experiments (orange solid lines), and finite element simulations (solid green lines). (d-f)~Schematic images describing the initial undeformed shapes of structures with one hole (white disks) and one inclusion (orange disks) in three different configurations (horizontal, vertical and inclined at $45^\circ$), and applied external strain $\varepsilon_{yy}^{\text{ext}}=-0.05$. Deformed contours of holes and inclusions obtained with elastic multipole method ($n_\text{max}=10$, blue dashed lines), experiments (orange solid lines), and finite element simulations (solid green lines).
  In all cases, the size of  samples was $100\text{ mm}\times100\text{ mm}\times25\text{ mm}$, the diameters of each hole/inclusion were $d=8.11 \text{ mm}$, and their  separation distances were $a=9.50 \text{ mm}$.
}
  \label{Fig:ExperimentalHolesC}
\end{figure}

We analyzed uniaxially compressed $100 \text{ mm}\times 100\text{ mm}\times 25\text{ mm}$ elastomer structures with three different configurations (horizontal, vertical and inclined at $45^\circ$ angle) of two holes with identical diameters $d=8.11$~mm and their separation distance  $a=9.50$~mm (see Fig.~\ref{Fig:ExperimentalHolesC}a-c). Holes were placed near the centers of elastomer structures to minimize the effects of boundaries. The structures were relatively thick ($25$~mm) to prevent the out-of-plane buckling. The contours of deformed holes in compressed experimental samples under external strain $\varepsilon_{yy}^{\text{ext}}=-0.05$ were compared with  those obtained with elastic multipole method and finite element simulations (see Fig.~\ref{Fig:ExperimentalHolesC}d-f). For the elastic multipole method, we used external stress $\sigma_{yy}^{\text{ext}}=E_0 \varepsilon_{yy}^{\text{ext}}$ ($\sigma_{xx}^{\text{ext}}\approx \sigma_{xy}^{\text{ext}}\approx 0$ due to reduced friction) and  plane stress condition was assumed  since the experimental samples were free to expand in the out-of-plane direction. Linear finite element simulations were performed for finite-size ($100 \text{ mm}\times100  \text{ mm}$) 2D structures  with circular holes under plane stress condition. In finite element simulations, samples were compressed by prescribing a uniform displacement in the $y$-direction on the upper and lower surfaces, while allowing nodes on these surfaces to move freely in the $x$-direction. The midpoint of the bottom edge was constrained to prevent rigid body translations in the $x$-direction.

\begin{figure}[!t]
\centering
  \includegraphics[width=\textwidth]{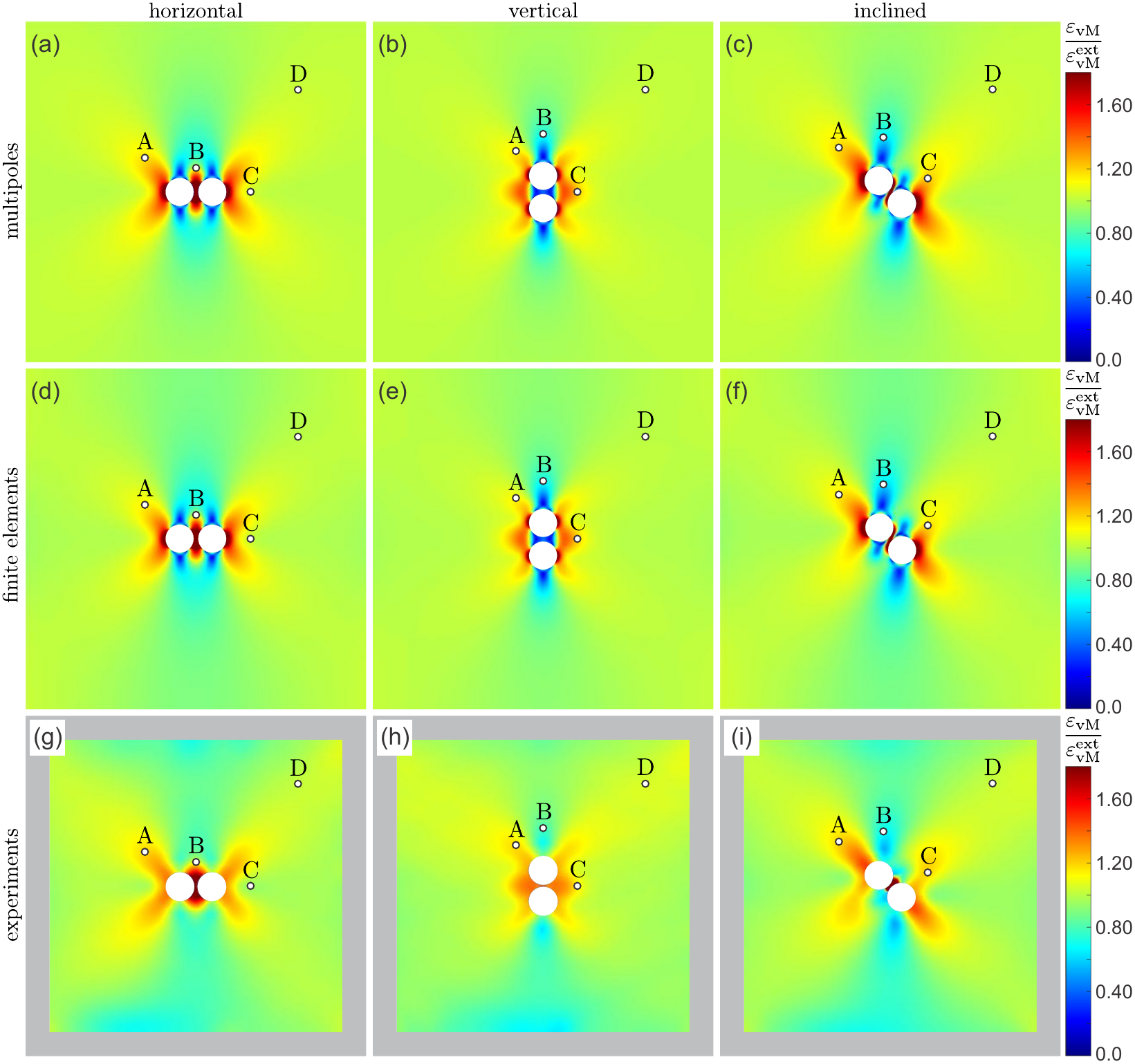}
  \caption{Equivalent von Mises strain fields $\varepsilon_\text{vM}$  (see Eq.~(\ref{eq:vonMisesStrain})) for uniaxially vertically compressed elastic structures with two holes (white disks) in three different configurations (horizontal, vertical, inclined at $45^\circ$) introduced in Fig.~\ref{Fig:ExperimentalHolesC}. Equivalent von Mises strain fields $\varepsilon_\text{vM}$ were obtained with (a-c)~elastic multipole method ($n_\text{max}=10$), (d-f)~finite element simulations, and (g-i)~DIC analysis of experiments. Note that the strain data was corrupted near the edges for some samples due to oil stains on the speckle patterns near the boundary. For this reason, we omitted the affected border regions (grey frames) in heat maps (g-i). Four marked points A-D were chosen for the quantitative comparison of strains $\varepsilon_\text{vM}$. See Table~\ref{tab:2holes} for details.}
  \label{Fig:ExperimentalHolesE}
\end{figure}

The contours of deformed holes obtained in experiments agree very well with those obtained with elastic multipole method ($n_\text{max}=10$) and linear finite element simulations for all three configurations of holes (see Fig.~\ref{Fig:ExperimentalHolesC}a-c). 
We also compared the equivalent von Mises strain fields defined as~\cite{Barber}
\begin{equation}
    \label{eq:vonMisesStrain}
  \varepsilon_{\text{vM}}=\frac{\sigma_\text{vM}}{E}=\frac{1}{1+\nu}\sqrt{ \varepsilon_{xx}^2-\varepsilon_{xx}\varepsilon_{yy}+\varepsilon_{yy}^2+3\varepsilon_{xy}^2+\frac{\nu}{(1-\nu)^2}(\varepsilon_{xx}+\varepsilon_{yy})^2}
\end{equation}
that were obtained with elastic multipole method ($n_\text{max}=10$), finite element simulations, and DIC analyses of experiments (see Fig.~\ref{Fig:ExperimentalHolesE}). For all three configurations of holes, the strain fields agree very well between the elastic multipole method (Fig.~\ref{Fig:ExperimentalHolesE}a-c) and finite element simulations (Fig.~\ref{Fig:ExperimentalHolesE}d-f). The strain fields for experimental samples are qualitatively similar, but they differ quantitatively near the holes as can be seen from heat maps in Fig.~\ref{Fig:ExperimentalHolesE}g-i. The quantitative comparison of strains at four different points A-D (marked in Fig.~\ref{Fig:ExperimentalHolesE}) showed a relative error of $2$--$4\%$ between elastic multipole method and finite elements, and a relative error of $0$--$14\%$ between elastic multipole method and experiments (see Table~\ref{tab:2holes}).
The discrepancy between elastic multipole method and finite element simulations is attributed to the finite size effects. For elastic multipole method, we assumed an infinite domain, while finite domains were modeled in finite element simulations to mimic experiments. Since the domains are relatively small, interactions of induced elastic multipoles with boundaries become important, which is discussed in detail in the companion paper~\cite{sarkar2020image}.
The discrepancy between experiments and elastic multipole method is also attributed to the confounding effects of nonlinear deformation due to moderately large compression ($\varepsilon_{yy}^{\text{ext}}=-0.05$), 3D deformation due to relatively thick samples, fabrication imperfections, nonzero friction between the sample and the mounting grips of the testing machine, the alignment of camera with the sample (2D DIC system was used), and the errors resulting from the choice of DIC parameters (see e.g. \cite{YuPan,Sutton}).

\begin{figure}[!t]
\centering
  \includegraphics[width=\textwidth]{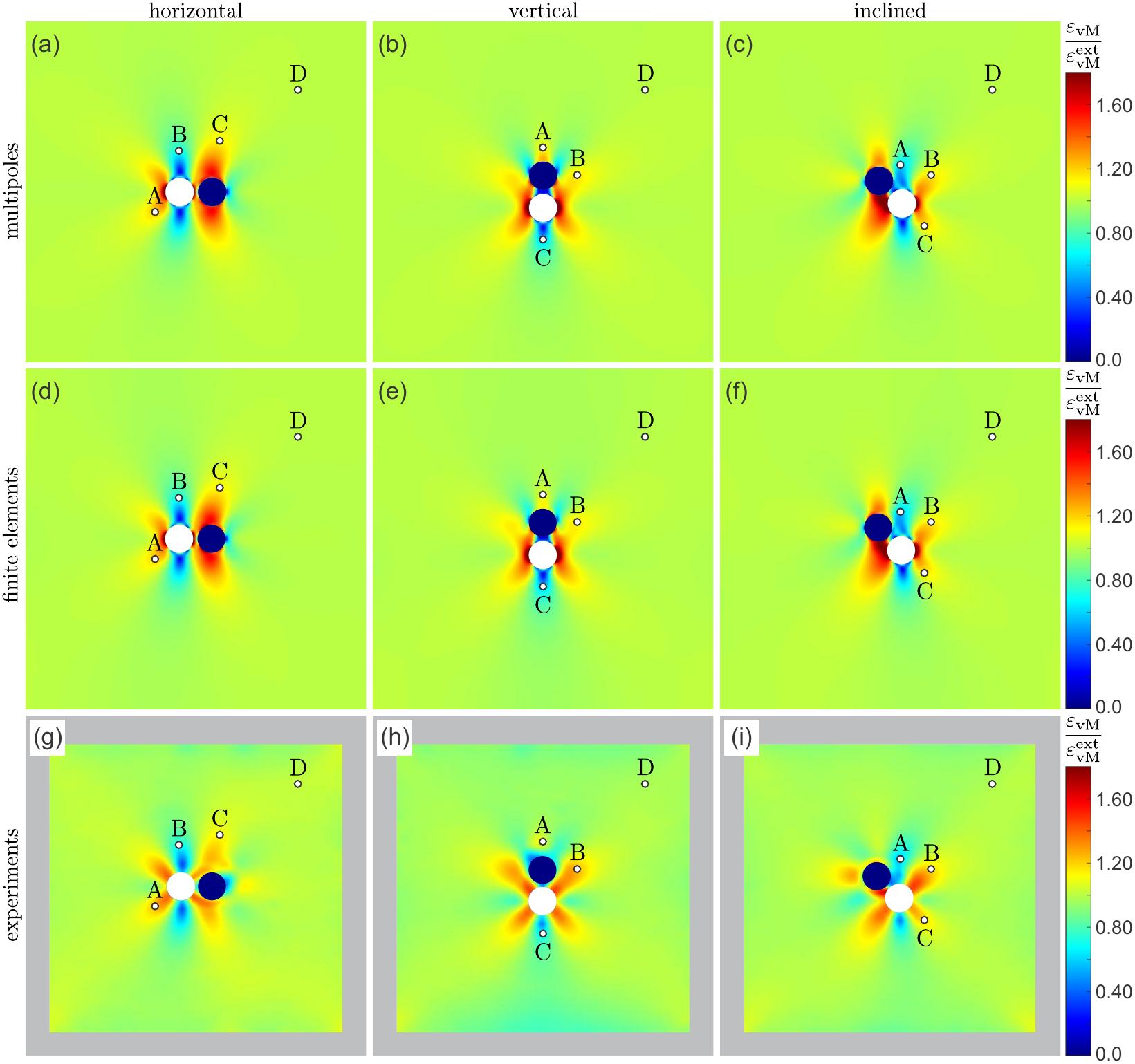}
  \caption{Equivalent von Mises strain fields $\varepsilon_\text{vM}$  (see Eq.~(\ref{eq:vonMisesStrain})) for uniaxially vertically compressed elastic structures with one hole (white disks) and one inclusion (blue disks) in three different configurations (horizontal, vertical, inclined at $45^\circ$) introduced in Fig.~\ref{Fig:ExperimentalHolesC}. Equivalent von Mises strain fields $\varepsilon_\text{vM}$ were obtained with (a-c)~elastic multipole method ($n_\text{max}=10$), (d-f)~linear finite element simulations, and (g-i)~DIC analysis of experiments. Note that the strain data was corrupted near the edges for some samples due to oil stains on the speckle patterns near the boundary. For this reason, we omitted the affected border regions (grey frames) in heat maps (g-i). Four marked points A-D were chosen for the quantitative comparison of strains $\varepsilon_\text{vM}$. See Table~\ref{tab:HoleInclusion} for details.}
  \label{Fig:ExperimentalInclusionsE}
\end{figure}

\begin{table}[!t]
\centering
\caption{Quantitative comparison for the values of equivalent von Mises strains $\varepsilon_\text{vM}$ normalized with the value for the applied external load $\varepsilon_\text{vM}^\text{ext}$ at points A-D (defined in Fig.~\ref{Fig:ExperimentalHolesE}) in compressed samples with two holes in three different configurations (horizontal, vertical, inclined) obtained with elastic multipole method (EMP), finite element simulations (FEM) and  DIC analysis of experiments (EXP). The relative percent errors between EMP and FEM were calculated as $100\times(\varepsilon_\text{vM}^\text{(EMP)}-\varepsilon_\text{vM}^\text{(FEM)})/\varepsilon_\text{vM}^\text{(FEM)}$. The relative percent errors between EMP and EXP were calculated as $100\times(\varepsilon_\text{vM}^\text{(EMP)}-\varepsilon_\text{vM}^\text{(EXP)})/\varepsilon_\text{vM}^\text{(EXP)}$.}
\label{tab:2holes}
\begin{tabular}{|c|ccc|>{\centering}p{0.07\textwidth}>{\centering}p{0.07\textwidth}|ccc|>{\centering}p{0.07\textwidth}>{\centering}p{0.07\textwidth}|ccc|>{\centering}p{0.07\textwidth}>{\centering\arraybackslash}p{0.07\textwidth}|}
\hline
\multirow{3}{*}{points}  & \multicolumn{5}{c|}{horizontal}& \multicolumn{5}{c|}{vertical}& \multicolumn{5}{c|}{inclined}\\
\cline{2-16}
& \multicolumn{3}{c|}{ strain $\varepsilon_\text{vM}/\epsilon_\text{vM}^\text{ext}$} & \multicolumn{2}{c|}{error of EMP (\%)} & \multicolumn{3}{c|}{ strain $\varepsilon_\text{vM}/\epsilon_\text{vM}^\text{ext}$} & \multicolumn{2}{c|}{error of EMP (\%)} & \multicolumn{3}{c|}{ strain $\varepsilon_\text{vM}/\epsilon_\text{vM}^\text{ext}$} & \multicolumn{2}{c|}{error of EMP (\%)} \\
\cline{2-16}
&  EMP   & FEM   & EXP   & FEM & EXP &  EMP   & FEM   & EXP   & FEM & EXP &   EMP   & FEM   & EXP   & FEM & EXP \\
\hline
A & 1.17 & 1.14 & 1.17 & 2.6 & 0.2  & 1.14 & 1.12 & 1.12 & 2.0 & 2.3  & 1.19 & 1.16 & 1.18 & 2.6 & 0.7 \\
B & 1.07 & 1.04 & 1.11 & 2.7 & 3.8  & 0.71 & 0.69 & 0.73 & 2.0 & 3.3  & 0.63 & 0.61 & 0.63 & 3.9 & 0.3 \\
C & 1.05 & 1.03 & 0.92 & 2.5 & 14.1 & 1.23 & 1.21 & 1.12 & 1.7 & 10.2 & 1.21 & 1.18 & 1.15 & 2.7 & 4.9 \\
D & 1.03 & 1.00 & 0.98 & 2.8 & 4.5  & 1.02 & 1.00 & 1.00 & 1.6 & 2.0  & 1.03 & 1.00 & 0.99 & 2.8 & 4.2 \\
\hline            
\end{tabular}
\end{table}

\begin{table}[!t]
\centering
\caption{Quantitative comparison for the values of equivalent von Mises strains $\varepsilon_\text{vM}$ normalized with the value for the applied external load $\varepsilon_\text{vM}^\text{ext}$ at points A-D (defined in Fig.~\ref{Fig:ExperimentalInclusionsE}) in compressed samples with one hole and one inclusion in three different configurations (horizontal, vertical, inclined) obtained with elastic multipole method (EMP), finite element simulations (FEM) and  DIC analysis of experiments (EXP). The relative percent errors between EMP and FEM were calculated as $100\times(\varepsilon_\text{vM}^\text{(EMP)}-\varepsilon_\text{vM}^\text{(FEM)})/\varepsilon_\text{vM}^\text{(FEM)}$. The relative percent errors between EMP and EXP were calculated as $100\times(\varepsilon_\text{vM}^\text{(EMP)}-\varepsilon_\text{vM}^\text{(EXP)})/\varepsilon_\text{vM}^\text{(EXP)}$.}
\label{tab:HoleInclusion}
\begin{tabular}{|c|ccc|>{\centering}p{0.07\textwidth}>{\centering}p{0.07\textwidth}|ccc|>{\centering}p{0.07\textwidth}>{\centering}p{0.07\textwidth}|ccc|>{\centering}p{0.07\textwidth}>{\centering\arraybackslash}p{0.07\textwidth}|}
\hline
\multirow{3}{*}{points}& \multicolumn{5}{c|}{horizontal}& \multicolumn{5}{c|}{vertical}& \multicolumn{5}{c|}{inclined}\\
\cline{2-16}
& \multicolumn{3}{c|}{ strain $\varepsilon_\text{vM}/\epsilon_\text{vM}^\text{ext}$} & \multicolumn{2}{c|}{error of EMP (\%)} & \multicolumn{3}{c|}{ strain $\varepsilon_\text{vM}/\epsilon_\text{vM}^\text{ext}$} & \multicolumn{2}{c|}{error of EMP (\%)} & \multicolumn{3}{c|}{ strain $\varepsilon_\text{vM}/\epsilon_\text{vM}^\text{ext}$} & \multicolumn{2}{c|}{error of EMP (\%)} \\
\cline{2-16}
&  EMP   & FEM   & EXP   & FEM & EXP &  EMP   & FEM   & EXP   & FEM & EXP &   EMP   & FEM   & EXP   & FEM & EXP \\
A & 1.21 & 1.20 & 1.23 & 0.7 & 1.1  & 1.11 & 1.07 & 1.06 & 3.9 & 4.5  & 0.61 & 0.59 & 0.57 & 4.8 & 7.0 \\
B & 0.70 & 0.70 & 0.62 & 0.6 & 11.9 & 1.14 & 1.14 & 1.27 & 0.5 & 10.1 & 1.18 & 1.15 & 1.18 & 2.4 & 0.5 \\
C & 1.14 & 1.13 & 1.09 & 0.7 & 4.0  & 0.57 & 0.56  & 0.61 & 1.4 & 6.3  & 1.16 & 1.15 & 1.17 & 0.7 & 1.2 \\
D & 1.01 & 1.00 & 0.97 & 0.2 & 3.7  & 1.00 & 0.99 & 0.97 & 1.0 & 3.3  & 1.00 & 0.99 & 0.96 & 0.8 & 4.0 \\
\hline            
\end{tabular}
\end{table}

Experiments were repeated with relatively rigid inclusions ($E_\text{inc}/E_0=3000$), where the samples described above were reused. Acrylic (PMMA) rods were inserted into one of the holes and glued with a cyanoacrylate adhesive for each of the  samples. The contours of deformed holes obtained in experiments matched very well with those obtained with elastic multipole method ($n_\text{max}=10$) and finite element simulations for all three configurations of holes and inclusions (see Fig.~\ref{Fig:ExperimentalHolesC}d-f). A relatively good agreement was also obtained for strain fields (see Fig.~\ref{Fig:ExperimentalInclusionsE}), where the strains inside rigid inclusions are very small (dark blue color). The quantitative comparison of strains at four different points A-D (marked in Fig.~\ref{Fig:ExperimentalInclusionsE}) showed a relative error of $0$-$5\%$ between elastic multipole method and finite elements, and a relative error of $0$-$12\%$ between elastic multipole method and experiments (see Table~\ref{tab:HoleInclusion}).

\section{Conclusion}\label{sec:Conclusion}

In this paper, we demonstrated how  induction and multipole expansion, which are common concepts in electrostatics, can be effectively used also for analyzing the linear deformation of infinite 2D elastic structures with circular holes and inclusions for both plane stress and plane strain conditions. Unlike in electrostatics, there are two different types of multipoles $\mbf{Q}_n^s$ and $\mbf{Q}_n^p$ in elasticity, which are derived from topological monopoles $s$ (disclinations) and non-topological monopoles $p$.  This is due to the biharmonic nature of the Airy stress function. The external load can induce all of these multipoles except for the topological defects called disclinations (topological monopole $s$) and dislocations (topological dipole $\mbf{d}^s$).

The multipole expansion is a so-called \textit{far-field} method and is thus extremely efficient when holes and inclusions are far apart. In this case, very accurate results can be obtained by considering only induced quadrupoles, because the effect of higher-order multipoles decays more rapidly at large distances. When holes and inclusions are closer together, their interactions via induced higher-order multipoles become important as well. The accuracy of the results increases exponentially with the maximum degree of elastic multipoles, which is also the case in electrostatics, and this is characteristic for spectral methods~\cite{SpecMethod}.

Note that the Stokes flows in 2D can also be described in terms of the biharmonic equations of the stream function~\cite{Stream}. Hence, it may seem that the multipole method described above could be adapted for Stokes flows around rigid and deformable obstacles. However, this is not possible due to the Stokes' paradox, which is the phenomenon that there can be no creeping flow of a fluid around a disk in 2D~\cite{Stream}.

The elastic multipole method presented here was limited to deformations of infinite structures with holes and inclusions of circular shapes. It can be generalized to deformations of finite size structures by employing the concept of image charges from electrostatics, which is discussed in detail in the companion paper~\cite{sarkar2020image}. This method can also be adapted to describe deformations of structures with non-circular holes and inclusions, and can in principle be generalized to describe deformations of curved thin shells with inclusions.

While the elastic multipole method presented here  focused only on  linear deformation, similar concepts can also be useful for describing the post-buckling deformation of mechanical metamaterials. Previously, it was demonstrated that the buckled patterns of structures with periodic arrays of holes~\cite{Kamien,matsumoto2012patterns,Auxetic,Moshe3}, square frames~\cite{Moshe4,Moshe5} and kirigami slits~\cite{Moshe4} can be qualitatively described with interacting quadrupoles. Furthermore, the approach with elastic quadrupoles has recently been extended to the  nonlinear regime of compressed structures with periodic arrays of holes, which can estimate the initial linear deformation, the critical buckling load, as well as the buckling mode~\cite{Moshe6}. The accuracy of these results could be further improved by expanding the induced fields to higher-order multipoles.  Thus, the elastic multipole method has the potential to significantly advance our understanding of deformation patterns in structures with holes and inclusions.

\section*{Acknowledgements}
This work was supported by  NSF through the Career Award DMR-1752100 and by the Slovenian Research Agency through the grant P2-0263. We would like to acknowledge useful discussions with Michael Moshe (Hebrew University) and thank Jonas Trojer (University of Ljubljana) for the help with experiments.
\appendix

\section{Linear finite element simulations}
\label{app:FEM}
Linear analyses in finite element simulations were performed with the commercial software Ansys\textsuperscript{\textregistered} Mechanical, Release 17.2. Geometric models of plates with holes and inclusions were discretized with 2D eight-node, quadratic elements of type PLANE183 set to the plane stress state option. The material for plates and inclusions was modeled as a linear isotropic elastic material. To minimize the effect of boundaries for the comparison with the elastic multipole method, which considers an infinite domain, we chose a sufficiently large square-shaped domain of size $L=400d$, where $d$ is the diameter of inclusions. To ensure high accuracy, we used a fine mesh with 360 quadratic elements evenly spaced around the circumference of each inclusion. 
To keep the total number of elements at a manageable number, the size of the elements increased at a rate of $2\%$ per element, when moving away from inclusions, up to the largest elements at the domain boundaries with an edge length of $L/200$. To prevent rigid body motions of the whole structure, we fixed the following 3 degrees of freedom: the displacement vector at the center of the square domain was specified to be zero ($u_x(0,0)=0$, $u_y(0,0)=0$); the midpoint of the left edge of the square domain was constrained to move only in the $x$-direction ($u_y(-L/2,0)=0$).
For consistency with finite element simulations, we imposed the same set of constraints ($u_x(0,0)=0$, $u_y(0,0)=0$, $u_y(-L/2,0)=0$) for the elastic multipole method. This was done in two steps. After obtaining the displacement field $(u_x^\text{EMP}(x,y),u_y^\text{EMP}(x,y))$
with the elastic multipole method, we first  subtracted the displacement $(u_x^\text{EMP}(0,0),u_y^\text{EMP}(0,0))$ at each point
\begin{subequations}
  \begin{align}
    {u'}_x^\text{EMP}(x,y) &= u_x^\text{EMP}(x,y)-u_x^\text{EMP}(0,0),\\
    {u'}_y^\text{EMP}(x,y) &= u_y^\text{EMP}(x,y)-u_y^\text{EMP}(0,0),
  \end{align}
\end{subequations}
to ensure that the center of the square domain is fixed (${u'}_x^\text{EMP}(0,0)={u'}_y^\text{EMP}(0,0)=0$). For this updated displacement field, the coordinates of points in the deformed configuration are $x'(x,y)=x+{u'}_x^\text{EMP}(x,y)$ and $y'(x,y)=y+{u'}_y^\text{EMP}(x,y)$. To impose the last constraint ($u_y(-L/2,0)=0$), this new deformed configuration was then rotated anticlockwise by the angle $\theta=\tan^{-1}({u'}_y^\text{EMP}(-L/2,0)/[L/2-{u'}_x^\text{EMP}(-L/2,0)])$ around the origin of the coordinate system as
\begin{subequations}
  \begin{align}
    x''(x,y)&= x'(x,y) \cos\theta - y'(x,y)  \sin\theta \equiv x + {u''}_x^\text{EMP}(x,y),\\
    y''(x,y)&=x'(x,y) \sin\theta + y'(x,y) \cos\theta \equiv y + {u''}_y^\text{EMP}(x,y).
  \end{align}
\end{subequations}
The set of displacement fields ${u''}_x^\text{EMP}(x,y)$ and ${u''}_y^\text{EMP}(x,y)$ was then used for comparison with finite element simulations.

\bibliographystyle{ieeetr}
\bibliography{refer.bib}

\end{document}